\documentclass{IEEEtran}
\usepackage{booktabs}
\usepackage{cite}
\usepackage{amssymb,amsfonts,amstext}% Lots of math symbols and enviro
\usepackage{amsmath}
\usepackage{graphicx} % For including graphics N.B. pdftex graphics dr
\usepackage{algorithm}
\usepackage{algorithmic}
\usepackage{multirow}
\usepackage[caption=false,font=footnotesize]{subfig}
\usepackage{url}
\usepackage{xcolor}
\usepackage{bbm}
\usepackage{psfrag}
\usepackage{enumitem}
\usepackage{diagbox}
\DeclareMathOperator*{\mini}{minimize}
\DeclareMathOperator*{\maxi}{maximize}
\DeclareMathOperator{\sbto}{subject \text{ } to}

\newtheorem{theorem}{Theorem}
\newtheorem{lemma}{Lemma}

\newtheorem{definition}{Definition}

\newcommand{\qed}{\nobreak \ifvmode \relax \else
      \ifdim\lastskip<1.5em \hskip-\lastskip
      \hskip1.5em plus0em minus0.5em \fi \nobreak
      \vrule height0.75em width0.5em depth0.25em\fi}

\newcommand{\bx}{\mathbf{x}}

\newcommand{\bd}{\mathbf{d}}

\newcommand{\bH}{\mathbf{H}}

\newcommand{\bW}{\mathbf{W}}

\newcommand{\Tr}{{\mbox{Tr}}}

\newcommand{\K}{{\cal K}}

\newcommand{\N}{{\cal N}}

\newcommand{\X}{{\cal X}}

\def\L{{\cal L}}

\ifCLASSINFOpdf
\else
\fi

\begin{document}
\title{Optimized Base-Station Cache Allocation for Cloud Radio Access Network 
with Multicast Backhaul}

\author{\IEEEauthorblockN{Binbin Dai, \IEEEmembership{Student Member, IEEE}, 
\IEEEauthorblockN{Ya-Feng Liu, \IEEEmembership{Member, IEEE}}, 
and Wei Yu, \IEEEmembership{Fellow, IEEE}}
%\IEEEauthorblockA{Department of Electrical and Computer Engineering\\
%         University of Toronto, Toronto, Ontario M5S 3G4, Canada  \\
%				Emails: \{bdai, weiyu\}@comm.utoronto.ca}
%\and
%\IEEEauthorblockN{Homer Simpson}
%\IEEEauthorblockA{Twentieth Century Fox\\
%\and
%\IEEEauthorblockN{James Kirk\\ and Montgomery Scott}
%\IEEEauthorblockA{Starfleet Academy\\
%San Francisco, California 96678-2391\\
%Fax: (888) 555--1212}

%\thanks{The materials in this paper have been presented in part at 
%the IEEE International Workshop on Signal Processing Advances in 
%Wireless Communications (SPAWC), Toronto, Canada, June 2014, \cite{binbin14}.
%This work was supported by Huawei Technologies Canada and by
%Natural Sciences and Engineering Research Council (NSERC) of Canada.
%The authors are with the Edward S. Rogers Sr. Department of Electrical
%and Computer Engineering, University of Toronto, Toronto, ON M5S 3G4, 
%Canada. (e-mails: bdai@ece.utoronto.ca, weiyu@comm.utoronto.ca).}

\thanks{Manuscript submitted on December 10, 2017; revised on April 11, 2018; accepted on April 18, 2018.
The materials in this paper have been presented in part at the 
IEEE International Conference on Acoustics, Speech and Signal Processing 
(ICASSP), Calgary, Canada, 2018 \cite{binbin18ICASSP}. 
This work was supported in part by the Natural Sciences and Engineering Research
Council (NSERC) of Canada and in part by the National Natural Science Foundation of China (NSFC) grants $11671419$ and $11688101$. 

B.~Dai and W.~Yu are with The Edward S. Rogers Sr. Department of Electrical
and Computer Engineering, University of Toronto, Toronto, ON M5S 3G4, 
Canada (e-mails: \{bdai, weiyu\}@comm.utoronto.ca). 
Y.-F.~Liu is with the State Key Laboratory of Scientific and Engineering Computing, 
Institute of Computational Mathematics and Scientific/Engineering Computing, Academy of 
Mathematics and Systems Science, Chinese Academy of Sciences, Beijing, 100190, China 
(e-mail: yafliu@lsec.cc.ac.cn). 
}
}


%Telephone: (800) 555--1212, Fax: (888) 555--1212}


\maketitle

\begin{abstract}

The performance of cloud radio access network (C-RAN) is limited by the finite
capacities of the backhaul links connecting the centralized processor (CP) with the base-stations
(BSs), especially when the backhaul is implemented in a wireless medium. 
%A promising approach to improving the performance of C-RAN is to augment the
%backhaul through BS caching, where the BSs pre-store contents of popular files.
This paper proposes the use of wireless multicast together with BS caching, where the BSs 
pre-store contents of popular files, to augment the backhaul of C-RAN.
For a downlink C-RAN consisting of a single cluster of BSs and wireless backhaul,
this paper studies the optimal cache size allocation strategy among the BSs and
the optimal multicast beamforming transmission strategy at the CP such that the user's
requested messages are delivered from the CP to the BSs in the most
efficient way. We first state a multicast backhaul rate
expression based on a joint cache-channel coding scheme, 
which implies that larger cache sizes should be allocated to the BSs 
with weaker channels. 
We then formulate a
two-timescale joint cache size allocation and beamforming design problem, where the
cache is optimized offline based on the long-term channel statistical
information, while the beamformer is designed during the file delivery phase 
based on the instantaneous channel state information. 
By leveraging the sample approximation method and the alternating
direction method of multipliers (ADMM), we develop efficient algorithms for optimizing
cache size allocation among the BSs, and quantify how much more cache  should be allocated 
to the weaker BSs.
We further consider the case with multiple files having different popularities
and show that it is in general not optimal to entirely cache the most
popular files first. Numerical results show considerable performance improvement of
the optimized cache size allocation scheme over the uniform allocation and other
heuristic schemes.

\end{abstract}


% \fi
\IEEEpeerreviewmaketitle

\section{Introduction} 

Cloud radio access network (C-RAN) has been recognized as one of the enabling
technologies to meet the ever-increasing demand for higher data rates for the 
next generation (5G) wireless networks \cite{Rost14, Simeone16, CRAN_book}. 
In C-RAN, the base-stations (BSs) are connected to a centralized processor
(CP) through high-speed fronthaul/backhaul links, which provide opportunities 
for cooperation among the BSs for inter-cell interference cancellation. The
performance of C-RAN depends crucially on the capacity of the
fronthaul/backhaul links. The objective of this paper is to explore the benefit
of utilizing caching at the BSs to augment the fronthaul/backhaul links. 

There are two fundamentally different fronthauling strategies that enable 
the cooperation of the BSs in C-RAN.  In the \emph{data-sharing} strategy 
\cite{simeone2009, marsch2008, Gesbert11,BinbinSparseBFJnal}, the CP
directly shares the user's messages with a cluster of BSs, which
subsequently serve the user through cooperative beamforming. In the
\emph{compression} strategy \cite{simeone2009, Park13}, the CP performs 
the beamforming operation and sends the compressed version of 
the analog beamformed signal to the BSs. 
%While the compression strategy has the computational benefit of offloading baseband processing to the cloud, it also requires higher fronthaul capacity to achieve good performance. 
The relative advantage of the data-sharing strategy versus the compression
strategy depends highly on the fronthaul/backhaul channel capacity
\cite{PratikEUSIPCO, Dai16}. In general, the compression strategy outperforms
data-sharing when the fronthaul/backhaul capacity is moderately high, in part because the
data-sharing strategy relies on the backhaul to carry each user's data multiple
times to multiple cooperating BSs. Thus, the finite backhaul capacity
significantly limits the BS cooperation size. 
%times, each time to a BS in the BS cluster, leading to a small BS cooperation size when the backhaul channel capacity is finite. 

%The data-sharing strategy typically performs better at low backhaul capacity, while the compression strategy is superior at moderate-to-high backhaul capacity. 

The capacity limitation in fronthaul/backhaul is especially pertinent for
small-cell deployment where high-speed fiber optic connections from the CP to
the BSs may not be available and {\it wireless backhauling} may be the most
feasible engineering option. The purpose of this paper is to point out that
under this scenario, the data-sharing strategy has a distinct edge in that it
can take advantage of: (i) the ability of the CP to {\it multicast} user
messages wirelessly to multiple BSs at the same time; and (ii) the ability of
the BSs to {\it cache} user messages to further alleviate the backhaul
requirement.  Note that the multicast opportunity in the wireless backhaul and the 
caching opportunity at the BSs are only available to facilitate the data-sharing
strategy in C-RAN, but not the compression strategy, as the latter involves
sending analog compressed beamformed signals from the CP to the BSs, which
are different for different BSs and are also constantly changing according to
the channel conditions, so are impossible to cache.  

%This paper considers the backhaul channel to be wireless shared medium, hence the cloud can share the user's clean data with all the BSs within the cluster simultaneously through multicasting.  In addition, we consider the potential of BS caching in reducing the wireless backhaul traffic.  By

This paper considers a downlink C-RAN in which the CP utilizes multiple
antennas to multicast user messages to a single cluster of BSs using the
data-sharing strategy, while the BSs pre-store fractions of popular contents
during the off-peak time and request the rest of the files from the CP using
coded delivery via the noisy wireless backhaul channel.  Given a total cache 
constraint, we investigate the optimal cache size allocation strategy across 
the BSs and the optimal multicast beamforming transmission strategy at the CP
so that the file requests can be delivered most efficiently from the CP to the
BSs. It is important to emphasize that the optimizations of the BS cache size 
allocation and the beamforming strategy at the CP occur in different timescales.  While the beamformer can dynamically adapt to the instantaneous
channel realization, the cache size is optimized only at the cache allocation
phase and can only adapt to the long-term statistics of the backhaul channel. %Therefore, we first devise the cache allocation scheme that only depends on backhaul channel statistics, then optimize the transmit beamformer for each channel realization in the file delivery phase with fixed cache size at each BS. 

This paper proposes a sample approximation approach to solve the above
two-timescale optimization problem. 
%The difficulty of the abovementioned two-time scale problem mainly lies on the cache allocation phase, even for the simplified single BS cluster setup considered in this paper. 
The optimal cache size allocation considers the long-term channel statistics 
in allocating larger cache sizes to the BSs with weaker backhaul channels, while
accounting for the potential effect of beamforming. It also considers the
difference in file popularities in caching larger portions 
of more popular files.

\subsection{Related Work}

While caching has been extensively used at the edge of Internet, the idea of
coded caching that takes advantage of multicast opportunity has recently
attracted extensive research interests due to the pioneering work of
\cite{FLC}, which uses the network coding method to simultaneously deliver
multiple files through a common noiseless channel to multiple receivers, each
caching different parts of the files. 
% A number of follow-up works have considered to extend the setup in \cite{FLC} to more practical scenarios including decentralized system \cite{AliDecentralized}, non-uniform user demands \cite{NiesenNonUniform, Ji17, Wang15}, multiple user requests at the same time \cite{JiTLC15a, Tandon17}, and heterogeneous cache sizes \cite{AmiriYG16b, Yener17}. 
This paper studies a different scenario in which the \emph{same} content is
requested by multiple receivers (BSs), hence no network coding is needed and
the coded multicasting in this paper refers to channel coding across multiple
wireless backhaul channels with \emph{different} channel conditions between the
CP transmitter and the BS receivers. 

%In particular, we consider a practical C-RAN model where each BS only caches a fraction of each file and the CP multicasts the rest to the BSs over the noisy wireless backhaul channel. 
%Under a total cache size constraint, we investigate the optimal strategy to distribute the cache size 
%among the BSs such that the user requests can be delivered to the BSs in the most efficient way. 

C-RAN with BS caching has been previously considered in \cite{Sezgin16, Tao16, Park16, Simeone17}, 
but most previous works assume fixed cache allocation among the BSs. 
More specifically, \cite{Sezgin16} and \cite{Tao16} examine how BS caching helps in reducing 
both backhaul capacity consumption and BS power consumption under given users' quality-of-service 
constraints; \cite{Park16} studies how BS caching changes the way that backhaul is utilized and proposes 
a similar scheme as in \cite{Patil14} that combines the data-sharing strategy 
and the compression strategy to 
improve the spectral efficiency of the downlink C-RAN. 
This paper differs from the above works in focusing on how to optimally allocate the cache sizes 
among the BSs and design multicast beamformers at the CP to improve the efficiency of 
sharing user's requested files via the wireless backhaul channel. 
%Note that the optimizations of cache size allocation and multicast beamformers happen in different time scales. 
%While the beamformers can be dynamically adapted to each channel realization, cache size allocation is only 
%optimized once at the cache deployment phase and can only adapt to the long-term statistics of the backhaul channel. 
%Therefore, we first devise a cache allocation scheme that only depends on channel statistics, then optimize the 
%transmit beamformer for each channel realization in the file delivery phase with fixed cache size at each BS. 

Previous works on caching strategy optimization rely on the assumptions of 
either simplified networks \cite{Gitzenis13, Shanmugam13} 
or Poisson distributed networks \cite{Debbah15, Cui16, Tao17} 
that are reasonable in a network with a large number of BSs and users, and
focus on analyzing how BS caching helps in improving the performance of the
BS-to-user layer.  This paper instead considers a C-RAN with a single cluster
of BSs and investigates how BS caching improves the efficiency of file delivery
between the cloud and the BSs layer.  

%An optimization framework based on sample approximation 
%method \cite{stochastic_book} and ADMM \cite{ADMM} is proposed to achieve an optimized cache allocation 
%strategy that outperforms simple heuristic algorithms. 

This paper is motivated by \cite{BidokhtiWT16} which shows from an
information-theoretical perspective the advantage of allocating
different cache sizes to different BSs depending on their channel conditions.
In addition, \cite{BidokhtiWT16} proposes a joint cache-channel coding scheme
that optimally utilizes the caches at the BSs in a broadcast erasure channel, 
which is further generalized to the degraded broadcast channel in \cite{BidokhtiWT17}.  
We take the findings in
\cite{BidokhtiWT16} one step further %in deriving a new multicast rate expression with BS caching and 
by considering the effect of multiple-antenna
beamforming in a downlink C-RAN backhaul network.  
We also extend \cite{BidokhtiWT16} to the case of multiple files with different
popularities and demonstrate that the optimal caching strategy also highly
depends on the file popularities. 

%We consider both the expected file downloading time and the expected file downloading rate, 
%averaged over the channel realizations, as the objective functions to 
%optimize the cache sizes allocated to each BS.
%We also consider the case with multiple files, which are requested with different probabilities, 
%and demonstrate that the optimal caching strategy also highly depends on the file popularities. 

\subsection{Main Contributions}

This paper considers the joint optimization of BS cache size allocation and
multicast beamformer at the CP in two timescales for a downlink C-RAN 
with a single BS cluster under the data-sharing strategy.  The main
contributions of this paper are summarized as follows:

\begin{itemize}

\item \emph{Problem Formulation}: We derive a new multicast backhaul rate
expression with BS caching based on the joint cache-channel coding scheme of 
\cite{BidokhtiWT16}. We then formulate two new cache size allocation problems of
minimizing the expected file downloading time and maximizing the expected file
downloading rate subject to the total cache size constraint.  The cache size 
allocation is optimized offline and is fixed during the file delivery phase, while
the transmit beamformers are adapted to the real-time channel realization. 

\item \emph{Algorithms}: We propose efficient algorithms for solving the
formulated cache size allocation problems. More specifically, to deal with the
intractability of taking expectation over the channel realizations in the
objective functions, we approximate the expectation via sampling
\cite{stochastic_book}. Note that the sample size generally needs 
to be large in order to guarantee the approximation accuracy. We further
propose to solve the sample approximation problem using the successive linear
approximation technique and the 
alternating direction method of multipliers (ADMM) algorithm \cite{ADMM}, which decomposes
the potentially large-scale problem (due to the large sample size) into many
small-scale problems on each sample. 

\item \emph{Engineering Insight}: 
We quantify how much cache should be allocated among the BSs in a practical 
C-RAN setup, and 
show that, as compared to
the uniform and proportional cache size allocation schemes, the proposed scheme
allocates aggressively larger cache sizes to the files with higher popularities, 
and for each
file the proposed scheme allocates aggressively larger cache sizes to the 
BSs with weaker backhaul channels. 

\end{itemize}

\subsection{Paper Organization and Notations}

The remainder of this paper is organized as follows. 
Section~\ref{sec:SystemModel} introduces the considered system model for C-RAN. 
We derive the backhaul multicast rate with BS caching in 
Section~\ref{sec:BC_w_caching} and 
state the problem setup in Section~\ref{sec:two_stage}. 
Sections~\ref{sec:single_file} and \ref{sec:multi_file} focus on 
the proposed cache size allocation schemes 
for the single file case and the multiple files case, respectively. 
Simulation results are provided in Section~\ref{sec:simulations}. 
Conclusions are drawn in Section~\ref{sec:conclusion}. 

Throughout this paper, lower-case letters (e.g. $x$) and lower-case bold letters (e.g. $\mathbf{x}$) denote scalars and column vectors, respectively. 
We use $\mathbb{C}$ to denote the complex domain. 
The transpose and conjugate transpose of a vector are denoted as $(\cdot)^{T}$ and $(\cdot)^H$, respectively. 
The expectation of a random variable is denoted as $\mathbb{E} \left [ \cdot \right]$. 
Calligraphy letters are used to denote sets.

%This demo file is intended to serve as a ``starter file''
%for IEEE conference papers produced under \LaTeX\ using
%IEEEtran.cls version 1.7 and later.
%% You must have at least 2 lines in the paragraph with the drop letter
%% (should never be an issue)
%I wish you the best of success.
%\hfill January 11, 2007
%
%\subsection{Subsection Heading Here}
%Subsection text here.
%
%
%\subsubsection{Subsubsection Heading Here}
%Subsubsection text here.

% An example of a floating figure using the graphicx package.
% Note that \label must occur AFTER (or within) \caption.
% For figures, \caption should occur after the \includegraphics.
% Note that IEEEtran v1.7 and later has special internal code that
% is designed to preserve the operation of \label within \caption
% even when the captionsoff option is in effect. However, because
% of issues like this, it may be the safest practice to put all your
% \label just after \caption rather than within \caption{}.
%
% Reminder: the "draftcls" or "draftclsnofoot", not "draft", class
% option should be used if it is desired that the figures are to be
% displayed while in draft mode.
%
%\begin{figure}[!t]
%\centering
%\includegraphics[width=2.5in]{myfigure}
% where an .eps filename suffix will be assumed under latex, 
% and a .pdf suffix will be assumed for pdflatex; or what has been declared
% via \DeclareGraphicsExtensions.
%\caption{Simulation Results}
%\label{fig_sim}
%\end{figure}

% Note that IEEE typically puts floats only at the top, even when this
% results in a large percentage of a column being occupied by floats.

\section{System  Model}\label{sec:SystemModel}

\begin{figure}[t]
	\psfrag{p}[tc][Bc][1]{Central Processor}
	
	\psfrag{a}[tc][cc][1.0]{Wireless}
	\psfrag{b}[tc][cc][1.0]{Backhaul}
	\psfrag{x}[tc][tc][0.8]{Cache}
	\psfrag{y}[tc][tc][0.8]{Cache}
	\psfrag{z}[tc][tc][0.8]{Cache}
	\psfrag{m}[tc][tc][0.9]{BS $1$}
	\psfrag{e}[tc][tc][0.9]{BS $2$}
	\psfrag{g}[tc][tc][0.9]{BS $3$}
	\psfrag{n}[tc][tc][0.9]{User}
	%\psfrag{f}[tc][tc][0.9]{User $2$}
  \centering
  \includegraphics[width= 0.45\textwidth]{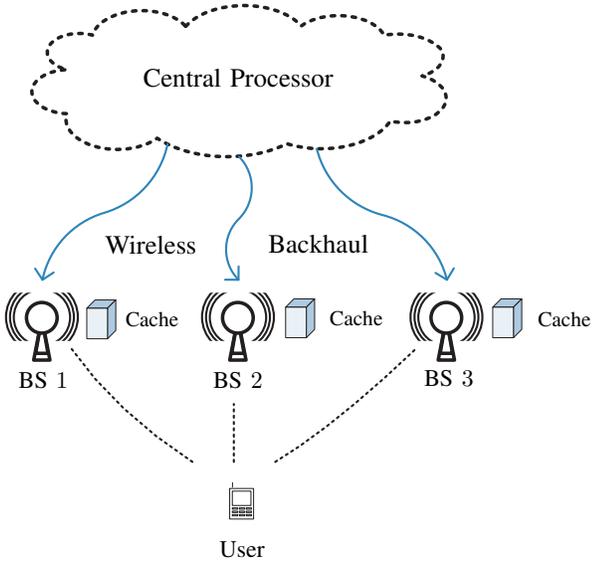}
\caption{Downlink C-RAN with BS caching, where 
each BS is equipped with a local storage unit that caches some contents of user's requested files.}
\label{fig:SystemModel}
\end{figure}

Consider a downlink C-RAN model in Fig.~\ref{fig:SystemModel} consisting of BSs
connected to a cloud-based CP through shared wireless backhaul. The cloud
employs a \emph{data-sharing} strategy which delivers each user's intended
message to a predefined cluster of BSs and the BS cluster subsequently serves
the user through cooperative beamforming.  The capacities of the backhaul is a
significant limiting factor to the performance of the C-RAN \cite{simeone2009,
BinbinSparseBFJnal}. 
To alleviate the backhaul requirement, %meet the demand for large BS cooperation size, 
this paper considers the scenario where each BS is equipped with a local cache, as shown in
Fig.~\ref{fig:SystemModel}, that can pre-store a subset of the files during
off-peak traffic time in order to reduce the peak time backhaul traffic.

For simplicity, we consider a network consisting of a single cluster of $L$
cooperative BSs, i.e., all the BSs in the network beamform cooperatively to
serve each user. In this case, the user's intended message needs to be made
available at all BSs in order to allow for cooperation.  We assume that the CP
has access to all the files and delivers the user's requested file to the BSs
through multicast beamforming over the wireless backhaul channel.   
%Suppose that there are a total of $K$ files in the library, where each file is of size 
%$F_k, k \in \K := \left\{1, 2, \ldots, K\right\}$.   

The backhaul connecting the CP with the BSs is implemented in a shared wireless
medium, assumed to follow a block-fading channel model.  We assume that the CP
is equipped with $M$ transmit antennas, while all the BSs are equipped with a
single antenna. We denote the channel vector between the CP and the BS
$l \in \mathcal{L} := \left\{1, 2, \ldots, L\right\}$, as
$\mathbf{h}_l \in \mathbb{C}^{M \times 1}$, which remains constant within a coherent
block but changes independently and identically according to some distribution
in different coherent blocks.  The received signal at BS $l$ can be written as
\begin{equation}\label{eqn:chn_model}
y_l = \mathbf{h}_l^{H} \mathbf{x} + z_l, 
\end{equation}
where $\mathbf{x} \in \mathbb{C}^{M \times 1}$ is the transmit signal
of the CP transmitter, $y_l \in \mathbb{C}$ is the received signal
at BS $l$, and $z_l \sim \mathcal{CN} \left(0,\, \sigma^2  \right)$ 
is the background noise at BS $l$ obeying the complex Gaussian distribution 
with zero mean and variance $\sigma^2$. 

Each BS $l$ is equipped with a local cache of size $C_l$ that can pre-store
some contents of the file. 
This paper addresses two questions:
\begin{itemize}
\item Given fixed local cache sizes and fixed cached contents, at a fast timescale, 
how should the transmit beamforming strategy be designed as function of the
instantaneous realization of the wireless channel in order to most efficiently
delivery a common user message to all the BSs? 
\item At a slow timescale, how should the contents be cached and how should the
cache sizes be allocated across the BSs so that the {\it expected} delivery
performance across many channel realizations is optimized? 
\end{itemize}
The answers to the above two questions would be 
trivial if the cache size at each BS is large enough 
to store the entire file library in the network, in which case no backhaul transmission is needed. 
This paper considers a more realistic scenario where the network operator has a fixed 
budget to deploy only a limited amount of total cache size $C$. 
Because of the 
limited cache size, each BS can only cache a subset of the files.  
In the next section, we define the file delivery performance in the backhaul link 
in terms of both the delivery rate and the downloading time, 
which are expressed as functions of cache sizes at the BSs.  

%In this paper, we address the question of how to distribute the total cache size $C$ among the BSs and design transmission strategies at the cloud such that the cloud can deliver the file to the BSs in the most effective way. 

%Two performance metrics are considered to evaluate the effectiveness of different caching strategies: 
%minimizing the expected file downloading time and maximizing the expected file downloading rate. 

\section{Broadcast Channel with Receiver Caching}\label{sec:BC_w_caching}

In this section, we investigate the optimal caching strategy for the backhaul
network with given cache size at each BS.  We then formulate the two-stage joint
cache and beamforming design problem considered in this paper in the next section.

\subsection{Separate Cache-Channel Coding}

Without BS caching, the downlink C-RAN wireless backhaul network with a single
cluster of BSs as shown in Fig.~\ref{fig:SystemModel} can be modeled as a
broadcast channel (BC) with common message only,
whose capacity is given as
\begin{equation} \label{eqn:r0}
R_0 \leq I\left( \mathbf{x} ; y_l \right), \forall ~l \in \L, 
\end{equation}
where $R_0$ denotes the multicast rate, ${I} \left( \mathbf{x} ; y_l \right)$ is
the mutual information between the transmit signal $\mathbf{x}$
at the CP and the received signal $y_l$ at BS $l$.
It can be seen from \eqref{eqn:r0} that the common information rate is limited
by the worst channel across the BSs.

To deal with the channel disparity issue in \eqref{eqn:r0}, 
this paper considers the use of BS caching
%, as opposed to  the use of secondary backhaul proposed in \cite{Liu17}, 
to smooth out the difference in channel quality across the BSs\footnote{In a similar vein, 
a related problem formulation of using secondary backhaul links to compensate for channel disparity is investigated in \cite{Liu17}.}. 
Assuming that BS $l$ has cache size $C_l$ bits, filled up by caching 
the first $C_l$ bits of a file with a total size of $F$ bits, a simple caching
strategy is to let the CP deliver only the rest $F - C_l$ bits of the file to
BS $l$. However, since the BSs are served through multicasting, the CP has
to send the maximum of the rest of the requested file, i.e., $\max_{l} \left\{
F - C_l \right\}$, to make sure that the BS with least cache size can get the
entire file.  Assuming that the channel coherent block is large enough so that
the file can be completely downloaded within one coherent block, then the
amount of time needed to finish the file downloading is
\begin{equation}\label{conventionalBC_time}
T_0 = \frac{\max_{l} \left\{ F - C_l \right\}}{\min_{l} \left\{{I} \left( \mathbf{x} ; y_l \right)\right\}}
\end{equation}
and the effective file downloading rate is  
\begin{equation}\label{conventionalBC}
D_0 = \frac{F}{T_0} =  \frac{\min_{l} \left\{{I} \left( \mathbf{x} ; y_l \right)\right\}}{\max_{l} \left\{ 1 - C_l/F \right\}}.
\end{equation}
As we can see from \eqref{conventionalBC_time} or \eqref{conventionalBC}, with
this naive caching strategy, it is optimal to allocate the cache
size uniformly among the BSs, i.e., $C_l = C/L, \forall~l\in\L$.

\subsection{Joint Cache-Channel Coding}

It is possible to significantly improve the naive separate cache-channel coding
strategy by considering cached content as side information for the broadcast
channel. The achievable rate of this strategy, named as joint cache-channel coding 
in \cite{BidokhtiWT16}, can be characterized as below.

\begin{lemma}[\cite{BidokhtiWT16}]\label{lemma:JCC}
Consider a BC with common message, 
if receiver $l \in\L$ caches $\alpha_l~(0 \leq \alpha_l \leq 1)$ fraction 
of the message, then the multicast common message rate $R$ is achievable 
if and only if the following set of inequalities are satisfied:
\begin{equation}\label{ineq:r_intra1}
R (1- \alpha_l) \leq {I} \left( \mathbf{x} ; y_l \right), \forall~l \in \mathcal{L}, 
\end{equation}
where ${\mathbf x}$ is the input and $y_l$'s are the output of the broadcast
channel.
\end{lemma}
\begin{IEEEproof}
We outline an information-theoretic proof as follows.
Consider that a message $w$ is chosen uniformly from the index set $\left\{ 1, 2, \ldots, W  \right\}$ and is to be 
transmitted to a set of receivers $\L$ over $n$ channel uses at a rate 
of $R = \frac{\log W}{n}$ bits per channel use. 
A codebook $\mathcal{C}$ of size $\left[ 2^{nR} , n\right]$ is first generated by drawing all symbols 
$x_i(j), i=1, 2, \ldots, n$, and $j = 1, 2, \ldots, 2^{nR}$, independently and identically according to the channel input distribution, 
where each row of $\mathcal{C}$ corresponds to a codeword. 
To send the message $w$, the $w$-th row of $\mathcal{C}$, denoted as 
$X^n(w) = \left[ x_1(w)~ x_2(w)~\ldots~x_n(w) \right]$, is transmitted over the channel. 
Note that the codebook $\mathcal{C}$ is revealed to both the transmitter and the $L$ receivers. 
After receiving $Y_l^n$, %the distorted version of $X^n(w)$ corrupted by the channel noise, 
the receiver $l$ tries to decode the index 
$w$ by looking for a codeword in the codebook $\mathcal{C}$ that is jointly 
typical with $Y_l^n$ and the cached content. 

Suppose that each receiver $l$ caches a fraction of the message---specifically caches the 
first $\alpha_l \log W$ bits of $w$.
Then, receiver $l$ only needs to search among those codewords whose indices 
start with the same $\alpha_l \log W$ bits 
as the cached bits. Since there exist a total number of $2^{nR - \alpha_l \log W}$ such codewords, 
by the packing lemma \cite{NIT12}, 
receiver $l$ would be able to find the correct codeword with diminishing 
error probability in the limit $n \rightarrow \infty$ as long as the inequality 
$nR - \alpha_l \log W \leq n I({\mathbf x}; y_l)$ is satisfied, or equivalently $R -
\alpha_l R \leq I({\mathbf x}; y_l)$, 
where $\mathbf{x}$ is the input channel symbol. 
This inequality needs to be satisfied by all $l\in\L$ 
to ensure that the common message is recovered by all the receivers, which leads to
the proof of the achievability of (\ref{ineq:r_intra1}) in Lemma~\ref{lemma:JCC}.
For the proof of converse, we refer to \cite{BidokhtiWT16} for the details. 
\end{IEEEproof}

%To apply the above lemma to the setup of this paper where the cache size $C_l$ is 
%defined in terms of bits as in the file size instead of bits/channel use as in 
%the file delivery rate, 
%we make an observation that if BS $l$ 
%caches $C_l / F$ fraction of the file, it can be equivalently seen as $C_l / F$ fraction of every 
%transmitted $R_c$ bits of the file are cached in BS $l$. Thus, BS $l$ only needs to have a channel 
%that can support decoding the rest $(1 - C_l / F) R_c$ bits. 
%Given this observation, the rate region in \eqref{ineq:r_intra1} can be rewritten as 
%\begin{equation}\label{ineq:r_signle_file}
%(1 - C_l / F) R_c \leq {I} \left( \mathbf{x} ; y_l \right), \forall~l \in \mathcal{L}.
%\end{equation}

In the setup of this paper, given cache size allocation $C_l$ and the file
size $F$, each BS $l$ can cache $C_l / F$ fraction of the file. 
Hence, by Lemma~\ref{lemma:JCC}, the file delivery rate $D_c$ 
with the joint cache-channel coding strategy can be formulated as 
\begin{equation}\label{ineq:r_intra2}
D_c = \min_l \left\{\frac{{I} \left( \mathbf{x} ; y_l \right)}{{1 - C_l/F}}\right\},
\end{equation}
and the downloading time $T_c$ can be written as 
\begin{equation}\label{ineq:r_intra2_time}
T_c = \frac{F}{D_c} = \max_l \left\{\frac{F - C_l}{{I} \left( \mathbf{x} ; y_l \right)}\right\}.
\end{equation}

Clearly, the above file downloading time and delivery rate are strictly better
than the ones in \eqref{conventionalBC_time} and \eqref{conventionalBC} except
when all ${I}\left( \mathbf{x} ; y_l \right)$ are equal to each other.  Instead
of allocating the cache size $C_l$ uniformly, \eqref{ineq:r_intra2} and
\eqref{ineq:r_intra2_time} suggest that it is advantageous to allocate more
cache to the BSs with weaker channels to achieve an overall higher multicast rate
or shorter downloading time. 
The difficulty, however, lies in the fact that in practice the channel condition 
changes over time while the cache size allocations 
among the BSs can only be optimized ahead of time at the cache 
deployment phase. 
In the next section, we formulate a two-stage optimization problem that jointly optimizes the 
cache size allocation strategy based on the long-term channel statistics 
and the beamforming strategy based on the short-term channel realization.

\section{Two-Stage Caching and Beamforming Design}\label{sec:two_stage}

We are now ready to formulate the two-stage joint cache size allocation and
beamforming design problem. At a slow timescale, cache size allocation is 
done at the cache deployment phase, so they can only
adapt to the channel statistics. At a fast timescale, the beamforming vector can
be designed to adapt to each channel realization during the file delivery
phase. 

First, we fix cache size allocation and content placement and focus on
the beamforming design in the fast timescale.  
%In a multiple-antenna BC, the
%transmit signal
Assuming that the BS uses a single-datastream multicast 
beamforming strategy for the multiple-antenna BC \eqref{eqn:chn_model}, the transmit signal is
%$\mathbf{x}$ in \eqref{eqn:chn_model} is formed by beamforming, as 
given by 
$\mathbf{x} = \mathbf{w} s$, where $\mathbf{w} \in \mathbb{C}^{M \times 1}$ is
the beamformer vector and $s \in \mathbb{C}$ is the user message, which can be
assumed to be complex Gaussian distributed $\mathcal{CN} \left(0,\, 1  \right)$. 
Then, the mutual information in the previous section becomes  
\begin{equation}
I({\mathbf x};y_l) =
\log \left( 1 + \frac{\text{Tr}\left( \mathbf{H}_l \mathbf{W}\right)}{\sigma^2}  \right), 
\end{equation}
where $\mathbf{H}_l = \mathbf{h}_l \mathbf{h}_l ^{H}$, and  
$\mathbf{W} = \mathbf{w}\mathbf{w}^H$ is the 
beamforming covariance matrix 
of the transmit signal $\mathbf{x}$ restricted within the constraint set 
\begin{equation}
\mathbb{W} = \left\{ \mathbf{W} \succeq \mathbf{0} \mid \text{Tr} \left(
\mathbf{W} \right) \leq P, \ \ \mathrm{rank}(\mathbf{W})=1  \right\}
\end{equation}  
with $P$ being the transmit power budget at the CP.

The above set is nonconvex due to the rank-one constraint. To obtain a numerical
solution, a common practice is to drop the rank-one constraint to enable convex
optimization, then to recover a feasible rank-one beamformer from the resulting
solution \cite{luo_davidson}.
While the solution so obtained is not necessarily global optimum, this strategy 
often works very well in practice, when compared to the globally 
optimal branch-and-bound algorithm \cite{Lu17}.

% Here, we relax the rank-one constraint for the covariance matrix $\mathbf{W}$ and assume that it can be of any rank. 

%The downloading time expression \eqref{Tc} suggests that the cache allocation $\left\{ C_l\right\}$ and the beamformer $\mathbf{W}$ can be jointly optimized when the cache sizes are subject to the total cache constraint $\sum_l C_l \leq C$. 

Under fixed channel realization ${\mathbf H}_l$ and cache size $C_l$, the optimal beamformer
design problem, after dropping the rank-one constraint, 
can now be formulated in terms of maximizing the delivery
rate (or equivalently minimizing the downloading time): 
\begin{subequations} \label{prob:bf_prob_orig}
\begin{align}
\displaystyle \maxi_{\left\{\mathbf{W}\right\}} \hspace{0.5mm} & \quad
\displaystyle  D_c \\
\sbto & \quad  \text{Tr} \left(  \mathbf{W} \right) \leq P,~\mathbf{W} \succeq \mathbf{0}, 
\end{align}
\end{subequations}
which can be reformulated as the following convex optimization problem: 
\begin{subequations} \label{prob:bf_prob}
\begin{align}
\displaystyle \maxi_{\left\{\mathbf{W},~\xi\right\}}  \hspace{0.5mm} & \quad
\displaystyle  \xi \\
 \sbto & \quad \log \left( 1+\frac{\Tr\left(\bH_l\bW\right)}{\sigma^2} \right) \geq \displaystyle \xi(F-C_l),  ~l\in\mathcal{L},  \\
 & \quad \text{Tr} \left(  \mathbf{W} \right) \leq P,~\mathbf{W} \succeq \mathbf{0}.
\end{align}
\end{subequations}
%The above beamforming design problem is equivalent to maximizing the downloading rate \eqref{ineq:r_intra2} under fixed cache 
%Problem \eqref{prob:bf_prob} is convex and 
This problem can be solved efficiently using standard optimization toolbox such as
CVX \cite{CVX}. To obtain a rank-one multicast beamforming vector afterwards,
we can adopt a strategy of using the eigenvector corresponding to the largest
eigenvalue of solution $\mathbf{W}^*$. The simulation section of this paper
later examines the performance loss due to such a relaxation of the rank-one
constraint. 

Next, we consider the allocation of cache sizes in the slow timescale.
The challenge is now to find the optimal allocation $C_l$ that 
minimizes the \emph{expected} file downloading time or maximizes the
\emph{expected} file downloading rate over the channel distribution. 
Intuitively, the role of caching at the BSs is to even out the channel
capacity disparity in the CP-to-BS links so as to improve the multicast rate,
which is the minimum capacity across the BSs. At the fast timescale,
transmit beamforming already does so to some extent. BS caching aims to
further improve the minimum. The challenge here is to optimize the cache size 
allocation, which is done in the slow timescale, while accounting for the
effect of beamforming, which is done in the fast timescale as a function
of the instantaneous channel.
We note that the caching strategy outlined in Lemma~\ref{lemma:JCC} is universal in 
the sense that it 
depends only on $C_l$ and not on ${\mathbf H}_l$. In the next
two sections, we devise efficient algorithms that optimize the 
cache size allocation at the BSs based on the long-term channel statistics using 
a sample approximation technique.

\section{Cache Allocation Optimization Across the BSs}
\label{sec:single_file}

In this section, we formulate the cache size allocation problem for delivering a
single file case of fixed size $F$ bits in order to illustrate a sample
approximation technique that allows us to quantify how much cache should
be allocated to the BSs with different average channel strengths.  
The multiple files case is treated in Section~\ref{sec:multi_file}. 

%Throughout this section, we assume that BS $l$ is equipped with a cache of size $C_l$ that is subject to the total cache constraint given by $\sum_l C_l \leq C$. 

\subsection{Minimizing Expected Downloading Time}

%Based on the channel model in \eqref{eqn:chn_model}, we substitute the 
%mutual information in the downloading time expression \eqref{ineq:r_intra2_time} with 
%$\log \left( 1 + \frac{\text{Tr}\left( \mathbf{H}_l \mathbf{W}\right)}{\sigma^2}  \right)$ and 
%formulate the optimal file downloading time  
%under the C-RAN setup of this paper with BS caching as 
%\begin{equation}\label{Tc}
%T_c^* = \min_{\mathbf{W} \in \mathbb{W}} \max_l\left\{
%\frac{F - C_l}{\log \left( 1 + \frac{\text{Tr}\left( \mathbf{H}_l \mathbf{W}\right)}{\sigma^2}  \right)}\right\},
%\end{equation} 
%in which $\mathbf{H}_l = \mathbf{h}_l \mathbf{h}_l ^{H}$, $\mathbf{W}$ is the covariance matrix 
%of the transmit signal $\mathbf{x}$ restricted within the constraint set 
%$\mathbb{W} = \left\{ \mathbf{W} \succeq \mathbf{0} \mid \text{Tr} \left(  \mathbf{W} \right) \leq P  \right\}$ 
%with $P$ being the transmit power budget at the cloud.
%Here, we do not restrict the covariance matrix $\mathbf{W}$ to be rank-one and assume that it can be of 
%any rank. 
%In the simulations, we will obtain a rank-one multicast beamforming vector by finding the 
%eigenvector corresponding to the largest eigenvalue of matrix $\mathbf{W}$ and examine the performance loss due to 
%the rank-one constraint. 
%
%Note that the optimal downloading time in \eqref{Tc} is a function of
%both the channel condition and the cache allocation. In practice,
%the transmit covariance matrix is adapted to each channel realization, but
%the cache allocation is determined at the cache deployment
%phase so it can only adapt to the channel statistics. 

For given cache size $C_l$, the optimal file downloading time
\eqref{ineq:r_intra2_time} can be written as 
\begin{equation}\label{Tc}
T_c^* = \min_{\mathbf{W}} \max_l\left\{
\frac{F - C_l}{\log \left( 1 + \frac{\text{Tr}\left( \mathbf{H}_l \mathbf{W}\right)}{\sigma^2}  \right)}\right\}.
\end{equation} 
Note that the file downloading time has also been considered in \cite{Simeone17, Xu17}
as the objective function.
Differently, in this paper, we take the expectation of $T_c^*$ over the channel distribution and aim to
find an optimal cache size allocation that minimizes the long-term expected file
downloading time. 
The cache optimization problem under a total cache size 
constraint $C$ across the BSs is formulated as:
\begin{subequations} \label{prob:mbf}
\begin{align}
 \mini_{\left\{ C_l \right\}}  \quad \hspace{1mm}& \mathbb{E}_{\left\{ \mathbf{H}_l \right\}} \left[T_c^*\right] \label{obj:mbf} \\
\sbto  \quad &  \sum_{l\in\L} C_l \leq C,~0 \leq C_l \leq F,~l  \in \mathcal{L}.
\end{align}
\end{subequations}
Finding a closed-form expression for the objective function in
\eqref{obj:mbf} is difficult. This paper proposes to
replace the objective function in \eqref{obj:mbf} with its sample
approximation \cite{stochastic_book} and to reformulate the problem as:
\begin{subequations} \label{prob:sample}
\begin{align}
\displaystyle \mini_{\left\{ C_l, ~\mathbf{W}^{n}\right\}}  \hspace{1mm} &
\quad  \frac{1}{N} \sum_{n=1}^{N} \max_l
\left\{\frac{F - C_l}{\log \left( 1 + \frac{\text{Tr}\left( \mathbf{H}_l^{n} \mathbf{W}^{n}\right)}{\sigma^2}  \right)}\right\} \label{obj:sample1}  \\
\sbto & \quad \sum_{l} C_l \leq C,~0 \leq C_l \leq F,~l \in \mathcal{L},  \label{total_cache_const} \\
 & \quad \text{Tr} \left(  \mathbf{W}^{n} \right) \leq P,~\mathbf{W}^{n} \succeq \mathbf{0},~ n \in \N, \label{power_const}
\end{align}
\end{subequations}
where $N$ is the sample size, $\N := \left\{ 1, 2, \ldots, N  \right\},$ $\left\{\mathbf{H}_l^n\right\}_{n\in\N}$ are the channel samples drawn according to the distribution of
$\mathbf{H}_l,$ and $\mathbf{W}^n$ is the beamforming covariance matrix
adapted to the samples $\left\{\mathbf{H}_l^n\right\}_{l\in\L}$.
%Note that in the simulations, we assume a particular channel distribution 
%in order to demonstrate the performance of the proposed algorithm. 
%However, the sample approximation method is applicable to any channel 
%distribution and is not restricted to the one considered in the simulations. 
Note that we do not assume any specific channel distribution here. In fact, 
the above sample approximation scheme works for any general channel distribution. 
Furthermore, even if in practice when the channel distribution is unknown, 
we can still use the historical channel realizations as the channel samples, as long as 
they are sampled from the same distribution.

Problem \eqref{prob:sample} is still not easy to solve mainly due to
the following two reasons.  First, the objective function of problem
\eqref{prob:sample} is nonsmooth and nonconvex, albeit all of its
constraints are convex.  Second, the sample size $N$ generally needs
to be sufficiently large such that the sample average is a good
approximation to the original expected downloading time, 
leading to a high complexity for solving problem \eqref{prob:sample}
directly.  In the following, we first reformulate problem
\eqref{prob:sample} as a smooth problem and linearize the nonconvex
term, then leverage the ADMM approach to decouple the problem
into $N$ low-complexity convex subproblems to improve the efficiency of 
solving the problem.

First, drop the constant $1/N$ in \eqref{obj:sample1} and introduce the auxiliary variable 
$\left\{\xi^n\right\}$, and reformulate problem \eqref{prob:sample} as
\begin{subequations} \label{prob:re_sample}
\begin{align}
\displaystyle \mini_{\left\{ C_l, ~\mathbf{W}^{n},~\xi^n\right\}}   & \quad
\displaystyle  \sum_{n=1}^N \frac{1}{\xi^n} \\
\sbto \hspace{1.2mm} & \quad \log \left( 1+\frac{\Tr\left(\bH_l^n\bW^n\right)}{\sigma^2}
\right) \geq \displaystyle \xi^n(F-C_l), \nonumber \\
%& \qquad \qquad \qquad \qquad ~l\in\mathcal{L},~n\in\N,   \label{const:nonconvex}\\
& \hspace{3.3cm} ~l\in\mathcal{L},~n\in\N  \label{const:nonconvex}, \\
& \quad \eqref{total_cache_const} ~ \text{and} ~ \eqref{power_const}. \nonumber
\end{align}
\end{subequations}

%{\color{blue}
%The above problem \eqref{prob:re_sample} is smooth but still nonconvex due to constraint \eqref{const:nonconvex}. 
%To see this, consider two points within \eqref{const:nonconvex}: 
%$\left( \left\{ C_l \right\}, \left\{ \mathbf{W}^n \right\}, \left\{ \xi^n \right\} \right) = 
%\left( \left\{ 0 \right\}, \left\{ \mathbf{0} \right\}, \left\{ 0 \right\}   \right)$ and 
%$\left( \left\{ F \right\}, \left\{ \mathbf{0} \right\}, \left\{ 1 \right\}   \right)$. 
%Now consider another point which is a convex combination of the previous two points: $\left( \left\{ F/2 \right\}, \left\{ \mathbf{0} \right\}, \left\{ 1/2 \right\}   \right)$. 
%This is clearly not in the constraint \eqref{const:nonconvex}.}
The above problem \eqref{prob:re_sample} is smooth but still nonconvex 
due to constraint \eqref{const:nonconvex}. 
To deal with this nonconvex constraint,
we approximate the nonconvex term $\xi^n(F-C_l)$ in \eqref{const:nonconvex} 
by its first-order Taylor expansion 
at some appropriate point $(\bar\xi^n, \bar C_l)$, i.e.,
\begin{align}
\xi^n(F-C_l) & \approx ~  \bar\xi^n (F-\bar C_l) + \left[F-\bar C_l, \, -\bar \xi^n \right] 
\cdot \nonumber \\
%& \qquad \qquad \qquad \qquad 
%\left[\xi^{n} - \bar \xi^n, \, {C_l} - {\bar C_l}\right]^{T} \\
& \hspace{3cm} \left[\xi^{n} - \bar \xi^n, \, {C_l} - {\bar C_l}\right]^{T}  \\
&  = ~ \bar\xi^{n} \left( F - C_l \right) + \left(F-\bar C_l\right) \left(\xi^n - \bar \xi^{n}\right). \label{eq:linear}
\end{align}

Based on \eqref{eq:linear}, an iterative first-order approximation is proposed in
Algorithm~\ref{alg:cache_alloc} for solving problem \eqref{prob:sample}. 
More specifically, let $\left\{\xi^n(t),\,C_l(t)\right\}$ be the iterates at the $t$-th iteration, the algorithm solves
\begin{subequations} \label{prob:re_sample2}
\begin{align}
\displaystyle \mini_{\left\{ C_l, ~\mathbf{W}^{n},~\xi^n\right\}}   & ~
\displaystyle  \sum_{n=1}^N \frac{1}{\xi^n} \\
\sbto \hspace{1.2mm} & ~ \log \left( 1+\frac{\Tr\left(\bH_l^n\bW^n\right)}{\sigma^2} \right) \geq \xi^{n}(t) \left( F - C_l \right)   \nonumber \\
& \hspace{0.3cm} + \left(F-C_l(t)\right) (\xi^n - \xi^{n}(t)), ~l\in\mathcal{L},~n\in\N, \label{lineart}\\
& \left \vert\xi^n - \xi^{n}(t) \right \vert \leq r(t), ~n\in\N, \label{radius_const}\\ 
& \left \vert C_l - C_l(t) \right \vert \leq r(t), ~l\in\L, \label{cache_radius_const}\\
&  \eqref{total_cache_const} ~ \text{and} ~ \eqref{power_const}, \nonumber
\end{align}
\end{subequations} 
with fixed $\left\{\xi^n(t),  C_l(t) \right\}$, 
where \eqref{radius_const} and \eqref{cache_radius_const} 
are the trust region constraints \cite{Nocedal2006NO}, within which we trust that 
the linear approximation in
\eqref{lineart} is accurate, 
and $r(t)$ is the trust region radius at the $t$-th iteration,  
which is chosen in a way such that the following condition is satisfied: 
\begin{equation}\label{reduction}
 \frac{\displaystyle \displaystyle\sum_{n=1}^N\frac{1}{\xi^{n}(t)}-  \sum_{n=1}^{N} \displaystyle \max_l
\left\{\frac{F - C_l^{\ast}(t)}{\log \left( 1 + \frac{\text{Tr}\left( \mathbf{H}_l^{n} \bW^{n\ast}(t)\right)}{\sigma^2}  \right)}\right\}}{\displaystyle \displaystyle\sum_{n=1}^N\frac{1}{\xi^{n}(t)}- \displaystyle\sum_{n=1}^N\frac{1}{\xi^{n\ast}(t)}}\geq \tau, 
\end{equation} 
where $\bW^{n\ast}(t),~C_l^{\ast}(t),~\xi^{n\ast}(t)$ are solutions to problem \eqref{prob:re_sample2} and $\tau\in(0,1)$ is a constant. 
Notice that the numerator in \eqref{reduction} is the actual reduction in the objective of problem \eqref{prob:sample} and 
the denominator is the predicted reduction. The condition in \eqref{reduction} basically says that the 
trust region radius is accepted only if the ratio of the actual reduction and the predicted reduction is greater than or equal 
to a constant, in which case problem \eqref{prob:re_sample2} is a good approximation of 
the original problem \eqref{prob:sample}.

\begin{algorithm}[t]
{\bf Initialization}: Initialize $C_l(1) = C/L,~l \in \mathcal{L},$ and $\xi^{n}(1)$ as the solution to problem
\eqref{prob:re_sample} with $C_l = C_l(1);$ set $t = 1$;\\
{\bf Repeat}: 
\begin{enumerate}
\item Initialize the trust region radius $r(t) = 1$; \\
{\bf Repeat}: 
\begin{enumerate}[leftmargin=20pt]
\item Use the ADMM approach in Appendix~\ref{apdx:a} to solve problem \eqref{prob:re_sample2}; 
\item Update $r(t) = r(t)/2$;   
\end{enumerate}
{\bf Until} condition \eqref{reduction} is satisfied.
%\item  %with fixed $\left\{\xi^n(t),  C_l(t) \right\}$ using the proposed ADMM approach in
%Section~\ref{sec:admm};
\item Update $\left\{\xi^n(t+1),  C_l(t+1)\right\}$ according to \eqref{update1} and \eqref{update2}, respectively;
\item Set $t = t + 1$;
\end{enumerate}
{\bf Until} convergence
\caption{Optimized Cache Allocation with Single File }
\label{alg:cache_alloc}
\end{algorithm}

After solving problem \eqref{prob:re_sample2}, the algorithm updates the parameters for the next iteration by substituting the solution obtained from the linearly 
approximated problem \eqref{prob:re_sample2} to the original problem \eqref{prob:sample}: 
\begin{align}
& \xi^{n}(t+1) = \displaystyle \min_{l\in\mathcal{L}} \left\{ \frac{\log \left( 1+\frac{\Tr\left(\bH_l^n\bW^{n\ast}(t)\right)}{\sigma^2} \right) }{F - C_l^{\ast}(t) }  \right\},~n\in\N, \label{update1}\\
& C_l(t+1) = C_l^{\ast}(t),~l\in\mathcal{L}. \label{update2}
\end{align}
%where $\bW^{n\ast}(t)$ and $C_l^{\ast}(t)$ are solutions to problem \eqref{prob:re_sample2}. 
For the initial point, we can set $C_{l}(1)$ to be ${C}/{L}$ for all $l\in\mathcal{L},$ then 
problem \eqref{prob:re_sample} can be decoupled 
into $N$ convex optimization subproblems
to solve for $\xi^{n}(1)$ for all $n\in\N$.

It remains to solve problem \eqref{prob:re_sample2}. Note that problem \eqref{prob:re_sample2} 
is a convex problem but with a potentially large number of variables due to the large sample size. 
We propose an ADMM approach \cite{ADMM} to solve problem \eqref{prob:re_sample2}, 
which decouples the high-dimensional problem into $N$ decoupled small-dimensional subproblems. 
The details of solving problem \eqref{prob:re_sample2} using the ADMM approach 
can be found in Appendix~\ref{apdx:a}. It can be shown that the ADMM approach is guaranteed 
to converge to the global optimum solution of the convex optimization problem \eqref{prob:re_sample2}.

Once problem \eqref{prob:mbf} is solved using Algorithm~\ref{alg:cache_alloc}, we 
fix the obtained optimized cache size allocation and 
evaluate its effectiveness under a different set of independently generated channels  
%than the $N$ sample channels used in problem \eqref{prob:sample}, 
and calculate the file downloading time\footnote{Note that the optimal downloading time 
for each given channel is the inverse of the optimal objective 
value of problem \eqref{prob:bf_prob}.} 
for each channel by solving the convex problem \eqref{prob:bf_prob}.

%In the final part of this section, 
%we prove the convergence of Algorithm~\ref{alg:cache_alloc}. 
Algorithm~\ref{alg:cache_alloc} is guaranteed to converge to a stationary point of 
the optimization problem \eqref{prob:sample}. In the rest of this section, we prove the convergence of Algorithm~\ref{alg:cache_alloc}.
First, we define the stationary point of problem \eqref{prob:sample} as in \cite{Yuan1985}.
\begin{definition}\label{def-stationary}
Consider a more general problem
\begin{equation}\label{generalproblem}\mini_{\bx\in\X} ~ F(\bx)\end{equation}
where $\X$ is the feasible set and $F(\bx)$ is defined as  
\begin{equation}\label{Fx} F(\bx):=\frac{1}{N} \sum_{n=1}^N \max_{l}\left\{ f_{nl}(\bx)\right\}.\end{equation}
Here, $\left\{ f_{nl}(\bx)\right\}$ is a set of 
continuously differentiable functions.  
Given any feasible point $\bar\bx,$ define
\begin{align}\label{stationarypoint}
   \Phi(\bar\bx) = 
	&\max_{
	\scriptsize
	\begin{array}{c}
\left\{ \|\bd\|\leq 1, \right. \\
\left. \bar\bx+\bd\in\X \right\}
\end{array}
}\left\{ \vphantom{\sum_{n=1}^N} F(\bar\bx) - \right. \nonumber \\
& \quad \quad ~ \left. \frac{1}{N} \sum_{n=1}^N \max_{l}\left\{ f_{nl}(\bar\bx) + \nabla f_{nl}(\bar\bx)^T \bd\right\}\right\}.  
\end{align} %and let $\bar \bd$ be the optimal solution of the above problem. 
A point $\bar \bx\in \cal X$ is called a stationary point of problem \eqref{generalproblem} if $\Phi(\bar \bx)=0.$ %in \eqref{stationarypoint} is equal to zero.
\end{definition}

Two remarks on the above definition of the stationary point are in order. 
First, it is simple to see that $\Phi(\bar \bx)$ is always nonnegative 
as $\bd=\mathbf{0}$ is a feasible point of  \eqref{stationarypoint}. If $\Phi(\bar \bx)=0,$ it means that there does not exist any feasible and decreasing direction at point $\bar\bx$ in the first-order approximation sense.  
Second, problem \eqref{prob:sample} is in the form of problem \eqref{generalproblem} 
if we set $\bx = \left\{C_l, \bW^n\right\},$ 
$f_{nl}(\bx)=\left(F - C_l\right) / \log \left( 1 + \frac{\text{Tr}\left( \mathbf{H}_l^{n} \mathbf{W}^{n}\right)}{\sigma^2}  \right),$ and $\X$ to be the feasible set of problem \eqref{prob:sample}, which is convex and bounded.

Based on the above stationary point definition, we now 
state the convergence result of Algorithm \ref{alg:cache_alloc} in the following theorem. 
\begin{theorem}\label{thm:conv}
Algorithm \ref{alg:cache_alloc} is guaranteed to converge. 
Any accumulation point of the sequence generated by Algorithm \ref{alg:cache_alloc} is a stationary point of problem \eqref{prob:sample}, or equivalently problem \eqref{prob:re_sample}.
\end{theorem}
\begin{IEEEproof}
Algorithm \ref{alg:cache_alloc} is a special case of the general nonsmooth trust region 
algorithm discussed in \cite[Chapter 11]{Conn2000}, which can be proved to converge 
to a stationary point of the general problem \eqref{generalproblem}. 
For completeness of this paper, we provide a proof outline 
in Appendix~\ref{convergence_proof}. 
\end{IEEEproof}

\subsection{Maximizing Expected Downloading Rate}

In this subsection, we consider maximizing the expected file downloading rate as the objective function to 
optimize the BS cache size allocation, which can be formulated as 
\begin{subequations} \label{prob:mbf_time}
\begin{align}
\displaystyle \maxi_{\left\{ C_l \right\}} \hspace{0.5mm}  &
\quad  \mathbb{E}_{\left\{ \mathbf{H}_l \right\}} \left[\frac{F}{T_c^*}\right] \label{obj:mbf_time} \\
\sbto & \quad \sum_{l\in\L} C_l \leq C,~0 \leq C_l \leq F,~l  \in \mathcal{L}, 
\end{align}
\end{subequations}
where $T_c^*$ is the optimal file downloading time defined in \eqref{Tc} 
under given channel realization and cache size allocation. 
Note that the expected value of the inverse of 
a random variable $X$, $\mathbb{E}\left[ \frac{1}{X} \right]$, 
is in general different from the inverse of the expected value of $X$, $\frac{1}{\mathbb{E}\left[ X \right]}$. 
Thus, the cache size allocation obtained from solving problem \eqref{prob:mbf_time}
is also different from 
the one obtained from solving problem \eqref{prob:mbf}. 

We use the same idea as in the previous subsection to solve problem \eqref{prob:mbf_time}. 
First, we replace the objective function \eqref{obj:mbf_time} with its sample approximation and reformulate the problem as 
\begin{subequations} \label{prob:re_sample_time}
\begin{align}
\displaystyle \maxi_{\left\{ C_l, ~\mathbf{W}^{n},~\xi^n\right\}}   & \quad
\displaystyle  \sum_{n=1}^N \xi^n \\
\sbto \hspace{1.2mm} & \quad \eqref{total_cache_const}, ~ \eqref{power_const}, ~ \text{and} ~  \eqref{const:nonconvex},  \nonumber
\end{align}
\end{subequations}
in which we have dropped the constants $N$ and $F$ from the objective function. 
Then, we replace the nonconvex term in constraint \eqref{const:nonconvex} by its linear approximation 
\eqref{eq:linear} and solve problem \eqref{prob:re_sample_time} via optimizing 
a sequence of linearly approximated problems similar to problem \eqref{prob:re_sample2}.  
The approximated problem at each iteration is solved via an ADMM approach similar to the one 
described in Appendix~\ref{apdx:a} with the only difference being that 
the first term $\frac{1}{\xi^n}$ in the subproblem \eqref{subproblem1} needs to be replaced by 
$-\xi^n$.  

Same as in the previous subsection, once the optimized cache size allocation is obtained from 
solving problem \eqref{prob:re_sample_time}, we evaluate its effectiveness on different sets of 
channels and solve the multicast rate\footnote{The optimal multicast rate for given channel is the optimal 
objective value of problem \eqref{prob:bf_prob}.}
by optimizing problem \eqref{prob:bf_prob}.

\section{Cache Allocation Optimization Across Files}
\label{sec:multi_file}

In this section, we consider the cache size allocation problem for the general case
with multiple files having different popularities.  Due to the minimal difference
between the downloading rate and the downloading time as described in the
previous section, we only focus on minimizing the expected file downloading
time as the objective function in this section. 

We assume that each file $k$ of equal size $F$ bits is requested from the user with given probability 
$p_k, k \in \K:=\left\{ 1, 2, \ldots, K  \right\}$, 
$\sum_k p_k = 1$, and that BS $l$ caches $C_{lk} / F$ fraction of file $k$ 
with a total cache size constraint given by 
$\sum_{l,k} C_{lk} \leq C$.  
Given that file $k$ is requested, according to Lemma~\ref{lemma:JCC}, 
the optimal downloading time for file $k$, denoted as $T_{k}^*$, can be written as 
\begin{equation}\label{Rc_lk}
T_{k}^* = \min_{\mathbf{W}_k \in \mathbb{W}} \max_l\left\{
\frac{F - C_{lk}}{\log \left( 1 + \frac{\text{Tr}\left( \mathbf{H}_l \mathbf{W}_k\right)}{\sigma^2}  \right)}\right\}. 
\end{equation}
Different from the downloading time \eqref{Tc} in the single file case, the above 
optimal downloading time $T_{k}^*$ is a random variable depending on not only the channel realization 
but also the index of the requested file. 
We take the expected value of $T_{k}^*$ on both the channel realization $\mathbf{H}_l$ and the file index $k$ as the 
objective function and formulate the multi-file cache size allocation problem as
\begin{subequations} \label{prob:mbf_multi}
\begin{align}
\displaystyle \mini_{\left\{ C_{lk} \right\}} \hspace{0.8mm}  &
\quad  \sum_{k} p_k \mathbb{E}_{\left\{ \mathbf{H}_l \right\}} \left[T_{k}^*\right] \label{obj:mbf_multi} \\
\sbto & \quad \sum_{l, k} C_{lk} \leq C,~0 \leq C_{lk} \leq F,~l  \in \mathcal{L}, ~ k \in \mathcal{K}. 
\end{align}
\end{subequations}

Although intuitively the more popular file should be allocated larger cache
size, the question of how much cache should be allocated to each file is
nontrivial. In particular, it is in general not true that one should allocate the most
popular file in its entirety first, then the second most popular file, etc., 
until the cache size is exhausted.
This is because the gain in term of the objective function of the optimization
problem \eqref{prob:mbf_multi} due to allocating progressively more cache size 
to one file diminishes as more cache is allocated. At some point, it is better 
to allocate some cache to the less popular files, even when the most popular
file has not been entirely cached. The optimal allocation needs to be found 
by solving problem \eqref{prob:mbf_multi}. 

To solve problem \eqref{prob:mbf_multi}, we use the same sample approximation idea as in the single file case. 
With an additional set of axillary variables $\xi_k^n$, problem \eqref{prob:mbf_multi} 
after the sample approximation can be formulated as 
\begin{subequations} \label{prob:re_sample_multi}
\begin{align}
\displaystyle \mini_{\left\{ C_{lk},~\mathbf{W}_k^{n},~\xi_k^n\right\}}   & \quad
\displaystyle  \sum_{k=1}^K p_k \sum_{n=1}^N \frac{1}{\xi_k^n} \\
\sbto \hspace{2.5mm} & \quad \log \left( 1+\frac{\Tr\left(\bH_{lk}^n\bW_k^n\right)}{\sigma^2} \right) \geq \displaystyle 
\xi_k^n(F-C_{lk}),  \nonumber \\ %~l\in\mathcal{L},~n\in\N,~k\in\K,  \label{const:nonconvex_multi}\\
& \hspace{2.1cm} ~l\in\mathcal{L},~n\in\N,~k\in\K  \label{const:nonconvex_multi}, \\
& \quad \sum_{l, k} C_{lk} \leq C,~0 \leq C_{lk} \leq F, \nonumber \\ 
& \hspace{3.3cm} ~ l \in \mathcal{L}, ~k \in \mathcal{K}, \label{total_cache_const_multi}\\
& \quad \text{Tr} \left(  \mathbf{W}_k^{n} \right) \leq P,~\mathbf{W}_k^{n}
\succeq \mathbf{0}, \nonumber \\ 
& \hspace{3.3cm}~ n \in \N, ~ k \in \K. \label{power_const_multi}
\end{align}
\end{subequations}
Problem \eqref{prob:re_sample_multi} is then solved in an iterative fashion. At each iteration 
the nonconvex term on the right hand side of \eqref{const:nonconvex_multi} is replaced  
by its first-order approximation and the resulting 
convex problem to be solved at the $t$-th iteration is given by:  
\begin{subequations} \label{prob:re_sample2_multi}
\begin{align}
\displaystyle \mini_{\left\{ C_{lk}, ~\mathbf{W}_k^{n},~\xi_k^n\right\}}   & ~
\displaystyle  \sum_{k=1}^K p_k \sum_{n=1}^N \frac{1}{\xi_k^n} \\
\sbto \hspace{2.5mm} & ~ \log \left( 1+\frac{\Tr\left(\bH_{lk}^n\bW_k^n\right)}{\sigma^2} \right) 
\geq \xi_k^{n}(t) \left( F - C_{lk} \right)   \nonumber \\
& \hspace{1cm}+ \left(F-C_{lk}(t)\right) (\xi_k^n - \xi_k^{n}(t)), \nonumber \\ 
& \hspace{2cm} ~ l\in\mathcal{L},~n\in\N,~k\in\K, \label{lineart_multi}\\
& \left \vert\xi_k^n - \xi_k^{n}(t) \right \vert \leq r(t), ~n\in\N,~k\in\K, \label{radius_const_multi}\\
& \left \vert C_{lk} - C_{lk}(t) \right \vert \leq r(t), ~l\in\mathcal{L},~k\in\K, \label{radius_const_multi_cache}\\
&  \eqref{total_cache_const_multi} ~ \text{and} ~ \eqref{power_const_multi}, \nonumber
\end{align}
\end{subequations} 
where $\xi_k^{n}(t)$ and $C_{lk}(t)$ are fixed parameters obtained from the previous iteration and are updated for the next 
iteration according to  
\begin{align}
& \xi_k^{n}(t+1) = \displaystyle \min_{l\in\mathcal{L}} \left\{ \frac{\log
\left( 1+\frac{\Tr\left(\bH_{lk}^n\bW_k^{n\ast}(t)\right)}{\sigma^2} \right)
}{F - C_{lk}^{\ast}(t) }  \right\}, \nonumber \\ 
& \hspace{3.8cm} ~ n\in\N,~k\in\K, \label{update1_multi}\\
& C_{lk}(t+1) = C_{lk}^{\ast}(t),~l\in\mathcal{L},~k\in\K, \label{update2_multi}
\end{align}
where $\bW_k^{n\ast}(t)$ and $C_{lk}^{\ast}(t)$ are solutions to problem \eqref{prob:re_sample2_multi}. 
Similar to \eqref{reduction}, the trust region radius $r(t)$ in \eqref{radius_const_multi} and 
\eqref{radius_const_multi_cache} is picked to satisfy the following condition: 
\begin{align}\label{reduction2}
& \frac{\displaystyle \displaystyle \sum_{k=1}^{K} p_k \sum_{n=1}^N \left( \frac{1}{\xi_k^{n}(t)}-  \displaystyle \max_l
\left\{\frac{F - C_{lk}^{\ast}(t)}{\log \left( 1 + \frac{\text{Tr}\left( \mathbf{H}_{lk}^{n} \bW_k^{n\ast}(t)\right)}{\sigma^2}  \right)}\right\} \right)}{ \displaystyle \sum_{k=1}^{K} p_k \sum_{n=1}^N\left(\frac{1}{\xi_k^{n}(t)}- \frac{1}{\xi_k^{n\ast}(t)} \right)} \nonumber \\
& \geq \tau 
\end{align} 
for some constant $\tau\in(0,1)$.

\begin{algorithm}[t]
{\bf Initialization}: Initialize $C_{lk}(1) = C/LK,~l \in \mathcal{L},~k\in\K$ and $\xi_k^{n}(1)$ as the solution to problem
\eqref{prob:re_sample_multi} with $C_{lk} = C_{lk}(1);$ set $t = 1$;\\
{\bf Repeat}:
\begin{enumerate}
\item Initialize the trust region radius $r(t) = 1$; \\
{\bf Repeat}: 
\begin{enumerate}[leftmargin=20pt]
\item Use the ADMM approach in Appendix~\ref{apdx:b} to solve problem \eqref{prob:re_sample2_multi}; 
\item Update $r(t) = r(t)/2$;   
\end{enumerate}
{\bf Until} condition \eqref{reduction2} is satisfied.
%\item Use the ADMM approach in Appendix~\ref{apdx:b} to solve problem \eqref{prob:re_sample2_multi}; %with fixed $\left\{\xi^n(t),  C_l(t) \right\}$ using the proposed ADMM approach in
%Section~\ref{sec:admm};
\item Update $\left\{\xi_k^n(t+1),  C_{lk}(t+1)\right\}$ according to \eqref{update1_multi} 
and \eqref{update2_multi}, respectively;
\item Set $t = t + 1$;
\end{enumerate}
{\bf Until} convergence
\caption{Optimized Cache Allocation with Multiple Files}
\label{alg:cache_alloc_multi}
\end{algorithm}

Note that problem \eqref{prob:re_sample2_multi} can also be solved by using an ADMM approach as 
explained in Appendix~\ref{apdx:b}, which decouples problem \eqref{prob:re_sample2_multi} into 
$NK$ subproblems and each subproblem corresponds to a pair of sample channel and file request. 
The overall proposed algorithm for solving the cache size allocation problem with multiple files 
is summarized in 
Algorithm~\ref{alg:cache_alloc_multi}. 
%, which converges to a stationary point of problem 
%\eqref{prob:mbf_multi} when the sample size $N$ is large enough. 

Once the cache size allocation for the multiple files case is optimized through Algorithm~\ref{alg:cache_alloc_multi}, 
we calculate the downloading time for file $k$ by solving problem \eqref{Rc_lk} with 
fixed $C_{lk}$, which can be formulated as a convex optimization problem similar to \eqref{prob:bf_prob}. 
We then compute the average downloading time by
averaging under different sets of channel realizations. %and compare the average downloading time with other caching schemes. 

%\subsection{Optimizing Cache Allocation via Minimizing Expected Downloading Time}

\section{Simulation Results}\label{sec:simulations}

\begin{figure}[t]
\centering
  \centering
  \includegraphics[width=0.5\textwidth]{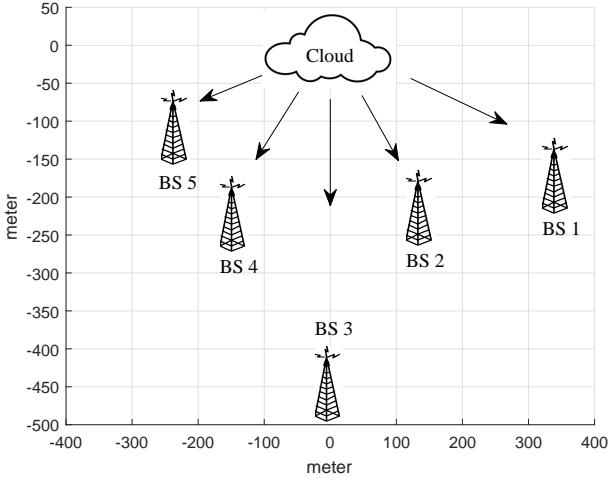}
\caption{A downlink C-RAN setup with $5$ BSs. The distances from the CP to the $5$ BSs are 
$(398, 278, 473, 286, 267)$ meters, respectively. }
\label{fig:topology}
\end{figure}

This section evaluates the performance of our proposed caching schemes through
simulations.  Consider a downlink C-RAN model with $L=5$ BSs randomly placed on
the half plane below the CP with the relative distances between the CP and the
$5$ BSs shown in Fig.~\ref{fig:topology}.  We generate $1000$ sets of channel
realizations from the CP to the BSs according to
$\mathbf{h}_l = \mathbf{K}_l^{1/2} \mathbf{v}_l$, where $\mathbf{K}_l$ models 
the correlation between the CP transmit antennas to BS $l$ and is 
generated mainly according to the angle-of-arrival and the antenna pattern, as
described in \cite{Lozano07}, with the path-loss
component modeled as $128.1 + 37.6\log_{10}(d)$ dB and $d$ is the distance
between the cloud and the BS in kilometers; $\mathbf{v}_l$ is a Gaussian random
vector with each element independently and identically distributed as
$\mathcal{CN} \left(0, 1  \right)$.  The first $N = 100$ sets of channel
realizations are used in the sample approximation to optimize the cache
allocation while the rest $900$ are used to evaluate the performance under the
obtained cache size allocation.  The details of the simulation parameters are listed
in Table~\ref{table:system-parameter}. 

\begin{table}[t]
\centering
\caption{Simulation Parameters.}
\label{table:system-parameter}
\begin{tabular}{|c|c|}
\hline 
Parameters & Values \\ \hline
Number of BSs & $5$ \\ \hline
Backhaul channel bandwidth & $20$ MHz    \\ \hline
Number of antennas at CP & $10$ \\ \hline
Number of antennas at each BS & $1$ \\ \hline
Maximum transmit power $P$ at CP &  $40$ Watts \\ \hline
 Antenna gain & $17$ dBi \\ \hline
 Background noise  & $-150$ dBm/Hz \\ \hline
 Path loss from CP to BS & $128.1+ 37.6 \log_{10}(d)$ \\ \hline
%Log-normal shadowing & $8$ dB \\ \hline
Rayleigh small scale fading & $0$ dB  \\ \hline
Normalized file size & $100$  \\ \hline
Training sample size $N$ & $100$ \\ \hline
Test sample size & $900$ \\ \hline
\end{tabular}
\end{table}

\subsection{Cache Allocation for BSs with Varying Channel Strengths}\label{sim:single_file}

In this subsection, we evaluate the performances of the proposed schemes 
for caching a single file across multiple BSs with different channel strengths
as discussed in Section~\ref{sec:single_file}. 
We compare the optimized cache size allocations obtained from minimizing the expected file downloading time \eqref{prob:mbf} 
and maximizing the expected file downloading rate \eqref{prob:mbf_time} with the following set of schemes: 

\begin{itemize}
\item \emph{No Cache}: Cache sizes $C_l = 0$ for all BSs;

\item \emph{Uniform Cache Allocation}: Cache sizes among the BSs are uniformly distributed as $C_l = C/L$, which serves as 
a baseline scheme; 

\item \emph{Proportional Cache Allocation}: Cache sizes among the BSs are proportionally allocated such that 
$\left(F - C_l\right) / \log \left( 1 + \frac{P\text{Tr}\left( \mathbf{K}_l \right)}{L\sigma^2}  \right)$ for all $l$ are equalized, if possible, which serves as another baseline scheme; 

\item \emph{Lower/Upper Bound}: 
Cache sizes among the BSs are dynamically and optimally allocated by solving
problem \eqref{prob:bf_prob} for each channel realization by treating
$\left\{C_l\right\}$ as the optimization variables, 
which is impractical in reality but serves as a 
lower bound for minimizing the expected file downloading time and an upper bound 
for maximizing the expected file downloading rate;
\item \emph{Rank-One Multicast Beamformer}: Cache sizes among the BSs are the same as the optimized caching schemes, but 
the multicast beamformer is restricted to be rank-one and is set to be the eigenvector corresponding to the largest eigenvalue 
of the optimized beamforming matrix $\mathbf{W}^n$ in each test sample channel. 
\end{itemize}

\begin{figure}[t]
\centering
  \centering
  \includegraphics[width=0.5\textwidth]{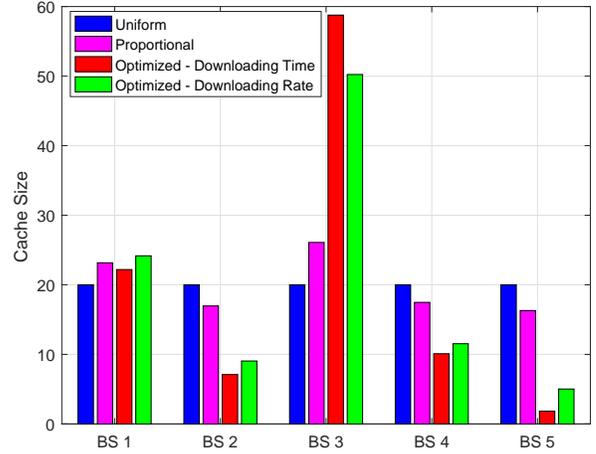}
\caption{Cache allocation for different schemes under total cache size $C = 100$, normalized 
with respect to file size $F=100$.  }
\label{fig:cache_bar}
\end{figure}

\begin{table*}[t]
\centering
\caption{File Downloading Time (ms/Mb) Comparison for Different Total Cache Sizes, 
Normalized with Respect to File Size $F=100$. 
%Without Caching, the Average Downloading Time is $11.45$ ms/Mb, 
%and the $90$th-Percentile Downloading Time is $14.76$ ms/Mb. 
}
\label{table:avg-downloading-time}
\begin{tabular}{|c|c|c|c|c|}
%\toprule
\hline
\multicolumn{1}{|c|}{\multirow{2}{*}{Cache Scheme} }  &  \multicolumn{2}{c}{Total Cache $C = 100$ } & \multicolumn{2}{|c|}{Total Cache $C = 200$ }\\  \cline{2-5}
\multicolumn{1}{|c|}{} & ~~~Average~~~ & $90$th-Percentile  & ~~~Average~~~  &  $90$th-Percentile   \\ \hline 
Uniform &  $9.15$       &  $11.8$  &  $6.87$    &   $8.86$  \\ \hline
Proportional & $8.63$   &  $10.91$      &    $6.47$   & $8.18$   \\ \hline
Optimized    & $7.68$   &  $8.54$       &   $5.76$   &  $6.42$    \\ \hline 
Rank-One     &  $7.85$  &  $8.69$       &   $5.86$   &  $6.51$     \\ \hline
Lower Bound  &  $6.89$  &  $7.56$       &   $5.13$   &  $5.59$     \\ \hline
%\bottomrule
\end{tabular}
\end{table*}

\begin{table*}[t]
\centering
\caption{File Downloading Rate (bps/Hz) Comparison for Different Total Cache Sizes, 
Normalized with Respect to File Size $F=100$. 
%Without Caching, the Average Downloading Rate is $4.63$ bps/Hz, 
%and the $10$th-Percentile Downloading Rate is $3.39$ bps/Hz.
}
\label{table:avg-downloading-rate}
\begin{tabular}{|c|c|c|c|c|}
%\toprule
\hline
\multicolumn{1}{|c|}{\multirow{2}{*}{Cache Scheme} }  &  \multicolumn{2}{c}{Total Cache $C = 100$ } & \multicolumn{2}{|c|}{Total Cache $C = 200$ }\\  \cline{2-5}
\multicolumn{1}{|c|}{} & ~~~Average~~~ & $10$th-Percentile  & ~~~Average~~~  &  $10$th-Percentile   \\ \hline
Uniform &  $5.79$      &  $4.24$  &  $7.71$    &   $5.65$  \\ \hline
Proportional & $6.11$    &   $4.58$    &  $8.15$   &   $6.11$   \\ \hline
Optimized    &  $6.64$    &   $5.78$    &  $8.85$     &   $7.71$  \\ \hline 
Rank-One     &   $6.58$   &  $5.65$     &   $8.78$     &   $7.54$  \\ \hline
Upper Bound  &   $7.29$    &  $6.62$    &   $9.78$     &  $8.95$   \\ \hline
%\bottomrule
\end{tabular}
\end{table*}

In Fig.~\ref{fig:cache_bar}, we compare the allocated BS cache sizes between the proposed schemes 
trained on the first $100$ channels and the 
baseline schemes under normalized file size $F=100$ and total cache size $C = 100$. 
As we can see, 
both of the proposed caching schemes are more aggressive in allocating larger cache sizes to the 
weaker BS $3$ as compared to the uniform and proportional caching schemes. 
We then evaluate the performances of different cache size allocation schemes 
on the rest $900$ sample channels 
and report the file downloading time and downloading rate (or spectral efficiency) 
in Table~\ref{table:avg-downloading-time} 
and \ref{table:avg-downloading-rate}, respectively, under two different
settings of total cache size $C=100$ and $C=200$, normalized with respect to 
file size $F=100$. 
As we can see, the proposed caching scheme improves over the uniform and 
proportional caching schemes by $10\% - 15\%$ on average,  but the gains are 
more significant for the $90$th-percentile downloading time and the $10$th-percentile 
downloading rate, which are around $20\% - 27\%$ and $26\% - 36\%$, respectively. 

We note here that without caching, the average and $90$th-percentile file 
downloading time are $11.45$ ms/Mb and $14.76$ ms/Mb, respectively, in this setting. 
The average and $10$th-percentile file downloading rate are $4.63$ bps/Hz and $3.39$ bps/Hz. 
Thus, the optimized BS caching schemes with $C=100$ and $C=200$ 
(normalized with respect to $F=100$) 
improve the average downloading time by about $33\%$ and $50\%$ respectively, and improve the average downloading rate by about $43\%$ and $91\%$ respectively.

\begin{figure}[t]
\centering
  \centering
  \includegraphics[width=0.45\textwidth]{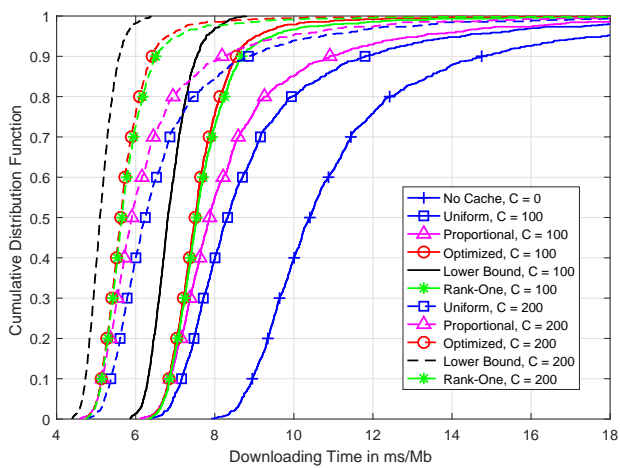}
\caption{CDF of downloading time under different caching schemes with total cache size 
$C=100$ and $C=200$, respectively, normalized with respect to file size $F=100$. }
\label{fig:cdf_time}
\end{figure}

\begin{figure}[t]
\centering
  \centering
  \includegraphics[width=0.45\textwidth]{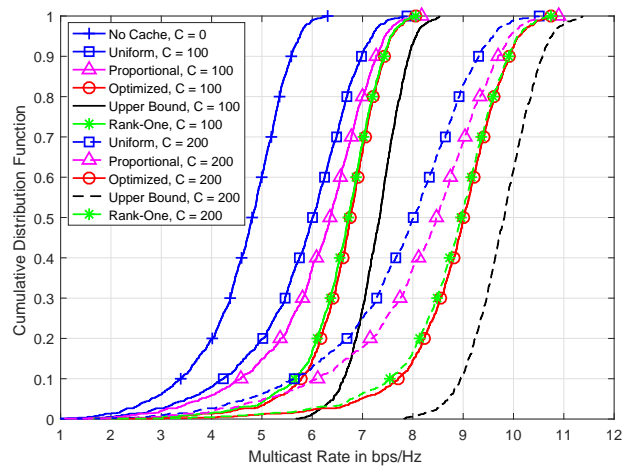}
\caption{CDF of downloading rates under different caching schemes with total cache size 
$C=100$ and $C=200$, respectively, normalized with respect to file size $F=100$. }
\label{fig:cdf_rate}
\end{figure}

In Figs.~\ref{fig:cdf_time} and \ref{fig:cdf_rate}, we compare the cumulative distribution functions (CDFs) of the 
downloading time and the downloading rates evaluated on the $900$ test channels with different caching schemes. 
Similar to what we have seen in Tables~\ref{table:avg-downloading-time} 
and \ref{table:avg-downloading-rate}, 
the proposed caching scheme shows significant gain on the high 
downloading time regime in Fig.~\ref{fig:cdf_time} 
and on the low downloading rate regime in Fig.~\ref{fig:cdf_rate} as compared to the baseline schemes. 
%For example, when $C=200$ bits, the proposed caching scheme by minimizing the expected file downloading time achieves a $90$-percentile 
%downloading time as $6.5$ ms/Mb in Fig.~\ref{fig:cdf_time} while the uniform and the proportional caching schemes require around $9$ ms/Mb and $8$ ms/Mb, respectively.  
%Similarly, in Fig.~\ref{fig:cdf_rate}, the proposed caching scheme by maximizing the expected file downloading rate achieves a 
%$10$-percentile downloading rate around $7.8$ bps/Hz when $C = 200$ bits as opposed to $5.5$ bps/Hz and $6$ bps/Hz for the uniform and the proportional 
%cache allocation schemes, respectively. 
From Figs.~\ref{fig:cdf_time} and \ref{fig:cdf_rate}, we can also see that the rank-one multicast beamformer shows negligible performance loss as compared to 
the general-rank multicast beamformer matrix $\mathbf{W}^n$ obtained by 
solving \eqref{prob:bf_prob}. 
It is also worth remarking that the lower bound scheme in Fig.~\ref{fig:cdf_time} and the upper bound scheme in Fig.~\ref{fig:cdf_rate} 
solve the cache size allocation problem dynamically for each channel realization, which is impractical, and only serve 
as benchmark schemes in this paper. 

To summarize the insight from the simulation results in this subsection for the single file case: 
First, although both the uniform and the proportional caching schemes perform fairly well in terms of the 
\emph{average} file downloading time and downloading rate, the proposed caching scheme shows significant 
gains in improving the high downloading time regime and the low downloading rate regime. 
This is due to the fact that BSs farther away from the cloud are more aggressively allocated larger amount 
of cache under the optimized scheme. 
Second, the rank-one beamformer derived from the general-rank covariance matrix does not degrade the 
performance much at all. Hence, we only focus on the performance of the proposed caching schemes 
without the rank-1 constraint on the covariance matrix in the next subsection for the multiple files case.

\subsection{Cache Allocation for Files of Varying Popularities}

\begin{table*}[t]
\centering
\captionsetup{justification=centering}
\caption{Optimized Cache Allocation $(C_{l1}, C_{l2})$ for a 2-File Case with Different File Popularities under $C=100$ and $F=100$.}
\label{table:cache-multi-file}
\begin{tabular}{|c|c|c|c|c|c|}
\hline 
%\multicolumn{1}{|c|}{\multirow{1}{*}{File}}  & \multicolumn{5}{c|}{} \\ \cline{2-6}
% \multicolumn{1}{|c|}{Popularity}     & $p_1 = 0.5, p_2 = 0.5$    &  $p_1=0.6, p_2=0.4)$  & $p_1=0.7, p_2=0.3$    &  $p_1=0.8, p_2=0.2$    &  $p_1=0.9, p_2=0.1$  \\ \hline
File  & \multirow{2}{*}{$(p_1, p_2) =(0.5, 0.5)$} & \multirow{2}{*}{$(p_1, p_2) =(0.6, 0.4)$} & \multirow{2}{*}{$(p_1, p_2) =(0.7, 0.3)$} & \multirow{2}{*}{$(p_1, p_2) =(0.8, 0.2)$} & \multirow{2}{*}{$(p_1, p_2) =(0.9, 0.1)$} \\
Popularity &           &         &       &     &     \\ \hline
BS1         & $(8.2, 8.2)$    &  $(13.2, 2.7)$   & $(16.8, 0)$     &  $(20.2, 0) $    &	 $(22.2, 0)$		       \\ \hline
BS2         & $(0, 0)$        &  $(0, 0)$        & $(0, 0)$        &  $(4.6, 0)$      &	 $(7.1, 0)$    \\ \hline
BS3         & $(41.8, 41.8)$  &  $(48.2, 35.9)$  & $(53.6, 27)$    &  $(56.8, 10.9)$  &	 $(58.8, 0)$      \\ \hline 
BS4         & $(0, 0)$        &  $(0, 0)$        & $(2.6, 0)$      &  $(7.5, 0)$      &	 $(10.1, 0)$       \\ \hline
BS5         & $(0, 0)$        &  $(0, 0)$        & $(0, 0)$        &  $(0, 0)$        &	 $(1.8, 0)$       \\ \hline
\hline
Total       & $(50, 50)$      &  $(61.4, 38.6)$  & $(73, 27)$      &  $(89.1, 10.9)$  &  $(100, 0)$ \\ \hline
\end{tabular}
\end{table*}

In this subsection, we present simulation results for the caching schemes with
multiple files having different popularities
and focus on the expected file downloading time as the performance metric. 
We first consider only two files with different pairs of request probabilities $(p_1, p_2)$ listed on the 
first row of Table~\ref{table:cache-multi-file}, where each column denotes the 
cache size allocation among the $5$ BSs under the specific file popularity given in the first row and each cell gives the 
cache size allocation between the two files within each BS. The cache sizes in
each column add up to the total cache size 
$C = 100$, normalized with respect to file size $F=100$. 

From Table~\ref{table:cache-multi-file} we see that for each column with given file popularity, 
the weakest BS $3$ always gets the most cache size as in the single file case shown 
in Fig.~\ref{fig:cache_bar}. 
Moreover, as the difference between the popularities of the two files increases across the columns, 
more cache is allocated to the first file. For example, the proposed caching scheme 
decides to allocate all the cache to only the more popular file $1$ when $(p_1, p_2) = (0.9, 0.1)$.

%\begin{figure}[t]
%\centering
  %\centering
  %\includegraphics[width=0.45\textwidth]{figs/mean_downloading_time_num_file.eps}
%\caption{Average downloading time for different numbers of files under the same Zipf distribution exponent $\alpha = 1$, where 
%the total cache size $C$ scales with the number of files $K$ as $100K$ bits. }
%\label{fig:mean_time_num_file}
%\end{figure}

\begin{figure}[t]
\centering
  \centering
  \includegraphics[width=0.45\textwidth]{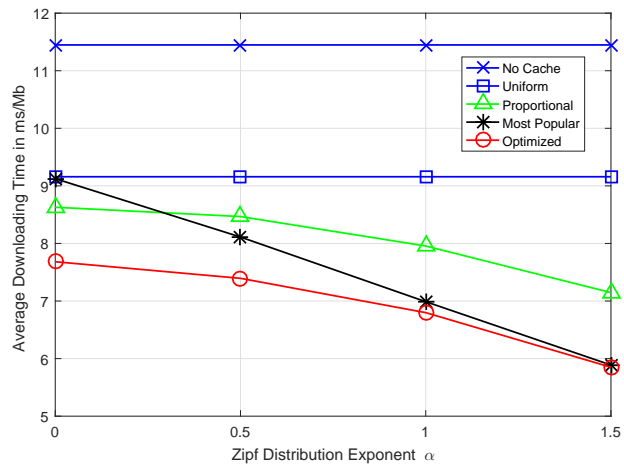}
\caption{Average downloading time for different Zipf file distributions under
the same number of files $K=4$ and total cache size $C=400$, normalized with respect 
to file size $F=100$. }
\label{fig:mean_time}
\end{figure}

In Fig.~\ref{fig:mean_time},  %and \ref{fig:mean_time_num_file}, 
we compare the average file downloading time between 
the optimized cache scheme and the following baseline schemes: 
\begin{itemize}
\item \emph{No Cache}: Cache size $C_{lk} = 0$ for all BSs and files; 

\item \emph{Uniform Cache Allocation}: Cache size for file $k$ at each BS $l$ is set to be as $C_{lk} = C/LK$ 
for all $k$ and $l$; 

\item \emph{Proportional Cache Allocation}: We first set the total cache size 
allocated for file $k$ as $p_kC$, then distribute $p_kC$ 
among the BSs according to the rule descried in the \emph{Proportional Cache Allocation} scheme in Section~\ref{sim:single_file}; 

\item \emph{Caching the Most Popular File}: We cache the most
popular file in its entirety first, then the second most popular file, etc. 
When a file cannot be cache entirely, we distribute the remaining cache among 
the BSs according to the \emph{Proportional Cache Allocation} scheme described in Section~\ref{sim:single_file}. 

\end{itemize}
In Fig.~\ref{fig:mean_time}, we fix the number of files to be $K=4$ and generate the file popularity 
according to the Zipf distribution \cite{Zink09} given by 
$p_k = \frac{k^{-\alpha}}{\sum_{i=1}^{K} i^{-\alpha}}, \forall~k,$ 
with different settings of $\alpha$. As the Zipf distribution exponent $\alpha$ increases, 
the difference among the file popularities also increases. 
As we can see from Fig.~\ref{fig:mean_time}, the average downloading time for all schemes, except 
for the uniform caching scheme, decreases as $\alpha$ increases. 
This is because in uniform cache allocation the cache size is the same for all files, hence  
the downloading time is the same no matter which file is requested. 
In contrast, all other three schemes tend to allocate more cache to the more popular files. 
In particular, the proposed caching scheme converges to the scheme of caching the most popular file when $\alpha = 1.5$, 
while it consistently outperforms the proportional caching scheme.

From Fig.~\ref{fig:mean_time} we conclude that
first, the uniform cache size allocation scheme performs poorly when the files have different popularities 
and especially when the difference is large.
Second, it is advantageous to allocate larger cache size to the more popular file, however, 
it is not trivial to decide how much more cache is needed for the more popular file. 
Our proposed caching scheme provides a better cache size allocation solution as compared to the 
heuristic proportional caching scheme and the most popular file caching scheme.

%
% Note that often IEEE papers with subfigures do not employ subfigure
% captions (using the optional argument to \subfloat), but instead will
% reference/describe all of them (a), (b), etc., within the main caption.

% An example of a floating table. Note that, for IEEE style tables, the 
% \caption command should come BEFORE the table. Table text will default to
% \footnotesize as IEEE normally uses this smaller font for tables.
% The \label must come after \caption as always.
%
%\begin{table}[!t]
%% increase table row spacing, adjust to taste
%\renewcommand{\arraystretch}{1.3}
% if using array.sty, it might be a good idea to tweak the value of
% \extrarowheight as needed to properly center the text within the cells
%\caption{An Example of a Table}
%\label{table_example}
%\centering
%% Some packages, such as MDW tools, offer better commands for making tables
%% than the plain LaTeX2e tabular which is used here.
%\begin{tabular}{|c||c|}
%\hline
%One & Two\\
%\hline
%Three & Four\\
%\hline
%\end{tabular}
%\end{table}

% Note that IEEE does not put floats in the very first column - or typically
% anywhere on the first page for that matter. Also, in-text middle ("here")
% positioning is not used. Most IEEE journals/conferences use top floats
% exclusively. Note that, LaTeX2e, unlike IEEE journals/conferences, places
% footnotes above bottom floats. This can be corrected via the \fnbelowfloat
% command of the stfloats package.

\section{Conclusion}\label{sec:conclusion}

This paper points out that caching can be used to even out the channel
disparity in a multicast scenario. We study the optimal BS cache size allocation
problem in the downlink C-RAN with wireless backhaul to illustrate the
advantage of multicast and caching for the data-sharing strategy.
We first derive the optimal
multicast rate with BS caching, then formulate the cache size optimization problem
under two objective functions, minimizing the expected file downloading time
and maximizing the expected file downloading rate, subject to the total cache
size constraint.  By leveraging the sample approximation method and ADMM,
we propose efficient cache size allocation algorithms that considerably outperform
the heuristic schemes.

%\balance


\appendices

\section{Proof of Theorem~\ref{thm:conv}}\label{convergence_proof}

 % $\bx(t)$ to denote the sequence generated by Algorithm \ref{alg:cache_alloc} at the $t$-th iteration.

%Now we are ready to present the convergence result. %  and  is a compact set. %the feasible set

%For succinctness, we give a proof outline.  
We use the notations 
introduced in Definition \ref{def-stationary}
in the following convergence proof.
First of all, it is simple to show that the objective sequence $\left\{F(\bx(t))\right\}$
generated by Algorithm~\ref{alg:cache_alloc}
monotonically decreases and is lower bounded by zero. % This further implies that the sequence $\left\{\text{Tr}\left( \mathbf{H}_l^{n} \mathbf{W}^{n}(t)\right)\right\}$
%  is uniformly bounded below.
Second, by using the continuously differentiable property of the function 
$f_{nl}(\bx)$, 
it can be shown that there always exists a trust region radius 
$r(t)$ such that the condition \eqref{reduction} is satisfied 
and that $r(t)$ is lower bounded by some constant $r>0$, i.e., $r(t) \geq r > 0$, for all $t$. 
Moreover, since the generated sequence $\left\{\bx(t)\right\}$ lies in the bounded set $\X$,
there must exist an accumulation point. 
Without loss of generality, let $\bar \bx$ denote an accumulation point of some convergent subsequence 
indexed by $\cal T$. 
%denoted by $\bar \bx$, which is also the convergence point of some 
%convergent subsequence of $\left\{\bx(t)\right\}$ indexed by the set $\cal T$, 
%denoted as $\left\{\bx(t)\right\}_{t\in\cal T}$. 
Finally, we show $\Phi(\bar \bx)=0$ by contradiction: 
Suppose that $\bar\bx$ is not a stationary point, i.e., $\Phi(\bar \bx)=\delta>0$, 
then there exists a subsequence of $\left\{\bx(t)\right\}_{t\in\cal T}$ that is 
sufficiently close to $\bar \bx$ such that 
\begin{equation} \label{contradiction}
\displaystyle \frac{1}{N}\displaystyle\sum_{n=1}^N\frac{1}{\xi^{n}(t)}- \frac{1}{N}\displaystyle\sum_{n=1}^N\frac{1}{\xi^{n\ast}(t)} \geq r \Phi(\bx(t))\geq \frac{r\delta}{2}, 
\end{equation}
where the first inequality is due to \cite[Lemma 2.1 (iv)]{AugmentedLA}. 
Combining \eqref{contradiction} with \eqref{reduction} and \eqref{update1}, we get  
%$$\frac{1}{N}\displaystyle\sum_{n=1}^N\frac{1}{\xi^{n}(t)}-\frac{1}{N}\displaystyle\sum_{n=1}^N\frac{1}{\xi^{n}(t+1)}\geq \frac{c\delta}{2}>0,$$
$$F(\bx(t))-F(\bx(t+1))\geq \frac{\tau r\delta}{2}>0,$$
which further implies that $F(\bx(t)) \rightarrow -\infty$ as $t\rightarrow +\infty$ in $\cal T$. This contradicts the fact that the sequence $\left\{F(\bx(t))\right\}$ is bounded below by zero. The proof is completed.

\section{The ADMM Approach to Solve Problem \eqref{prob:re_sample2}}\label{apdx:a}

To apply the ADMM approach to solve problem \eqref{prob:re_sample2}, we first introduce a set of so-called 
consensus constraints $C_l^n=C_l,~l\in\mathcal{L},~n\in\N$, and reformulate problem \eqref{prob:re_sample2} as
\begin{subequations} \label{prob:re_sample3}
\begin{align}
\displaystyle \mini_{
\scriptsize
\begin{array}{c}
\left\{ \xi^n, \bW^n, \right. \\
\left. C_l^n, C_l\right\}
\end{array}
} & ~
\displaystyle  \sum_{n=1}^N \frac{1}{\xi^n}  \label{obj:new_admm}\\
\sbto & ~ \log \left( 1+\frac{\Tr\left(\bH_l^n\bW^n\right)}{\sigma^2} \right) \geq \xi^{n}(t) \left( F - C_l^n \right)   \nonumber \\
& \hspace{0.3cm} + \left(F - C_l(t) \right)\left(\xi^n - \xi^{n}(t)\right),~l\in\L,~n\in\N,  \label{const: admm1} \\
& C_l^n = C_l,~l\in\mathcal{L},~n\in\N, \label{const:consensus}\\
& \left \vert \xi^n - \xi^{n}(t) \right \vert \leq r(t), ~n\in\N,  \label{const: admm5} \\
& \left \vert C_l - C_l(t) \right \vert \leq r(t), ~l\in\L, \label{const: cache_radius}\\
&  \eqref{total_cache_const} ~ \text{and} ~ \eqref{power_const}, \nonumber
\end{align}
\end{subequations}
where we replace the variable $C_l$ in \eqref{lineart} with the newly introduced variable $C_l^n$ in \eqref{const: admm1}. 
We form the partial augmented Lagrangian of problem \eqref{prob:re_sample3} 
by moving the constraint \eqref{const:consensus} to the objective function \eqref{obj:new_admm} as follows:  
\begin{align}\label{AL}
 & \mathcal{L}_{\rho}\left( \xi^n, \bW^n,C_l^n, C_l; \lambda_{l}^n \right)   = \sum_{n=1}^N \frac{1}{\xi^n} + \\ \nonumber 
 & \hspace{1cm}  \sum_{l\in\mathcal{L}}\sum_{n\in\N} \left[\lambda_{l}^n\left(C_l^n-C_l\right)+\frac{\rho}{2}\left(C_l^n-C_l\right)^2\right],
\end{align}
where $\lambda_l^n$ is the Lagrange multiplier corresponding to the constraint $C_l=C_l^n$ and $\rho>0$ is the penalty parameter.

The idea of using the ADMM approach to solve \eqref{prob:re_sample3} is to 
sequentially update the primal variables via minimizing 
the augmented Lagrangian \eqref{AL}, followed by an update of the Lagrange multiplier. 
Particularly, at iteration $j+1$, the ADMM algorithm updates the variables 
according to the following three steps: 
\begin{itemize}[leftmargin=0.5in]
  \item [Step 1] Fix $\left\{C_l, \lambda_l^n\right\}^j$ obtained from iteration $j$, update $\left\{\xi^n, \bW^n, C_l^n\right\}$ for iteration $j+1$ as the solution to the following problem
	\begin{align*}
	 \mini_{\left\{\xi^n, \bW^n, C_l^n\right\}} & \quad \mathcal{L}_{\rho}\left( \xi^n, \bW^n,C_l^n, \left\{C_l\right\}^j; \left\{\lambda_l^n\right\}^j \right) \\
	 \sbto \hspace{0.5mm} & \quad \eqref{const: admm1}, ~ \eqref{const: admm5}, ~\text{and} ~\eqref{power_const}. \nonumber 
\end{align*}
	
  \item [Step 2] Fix $\left\{\xi^n, \bW^n,C_l^n \right\}^{j+1}$ obtained from Step 1, update 
	$\left\{C_l\right\}$ for iteration $j+1$ as the solution to the following problem 
	\begin{align*} 
	\mini_{\left\{C_l\right\}} \hspace{1mm} & \quad \mathcal{L}_{\rho}\left( \left\{\xi^n, \bW^n,C_l^n \right\}^{j+1}, C_l; \left\{\lambda_l^n\right\}^j \right) \\
	 \sbto  & \quad \eqref{const: cache_radius}, ~ \eqref{total_cache_const}. \nonumber 
\end{align*}
	
  \item [Step 3] Fix $\left\{ C_l^n \right\}^{j+1}$ and$\left\{ C_l\right\}^{j+1}$ 
	obtained from Steps 1 and 2 respectively, update the 
	Lagrange multiplier as:
	\begin{equation*} 
	\left\{\lambda_l^n\right\}^{j+1} = \left\{\lambda_l^n\right\}^j + 
	\rho\left(\left\{ C_l^n \right\}^{j+1}- \left\{C_l\right\}^{j+1}\right) . 
	\end{equation*}
\end{itemize}

In the above Step 1, the optimization problem is decoupled among the channel realizations and for each channel realization $n \in \N$ we 
solve the following subproblem:
\begin{subequations}\label{subproblem1}
 \begin{align} 
\displaystyle \mini_{\left\{ \xi^n, \bW^n,C_l^n\right\}} \quad &  \displaystyle  \frac{1}{\xi^n} +  \sum_{l\in\L} \left[\lambda_{l}^{n}\left(C_l^n-C_l\right)+\frac{\rho}{2}\left(C_l^n-C_l\right)^2\right]\\
 \sbto \hspace{1mm} \quad &  \log \left( 1+\frac{\Tr\left(\bH_l^n\bW^n\right)}{\sigma^2} \right) \geq \xi^{n}(t) \left( F - C_l^n \right)  \nonumber \\ 
& \hspace{0.3cm} + (F-C_l(t)) \left(\xi^n - \xi^{n}(t)\right), ~l\in\L,  \\ 
& \Tr(\bW^n) \leq P, ~\bW^n \succeq\mathbf{0},\\ 
& \left \vert \xi^n - \xi^{n}(t) \right \vert \leq r(t),% ~n\in\N
\end{align} 
\end{subequations}
where $C_l$ and $\lambda_{l}^n$ are fixed constants obtained from the previous iteration and set to be as 
$C_l = C_l^{j}, \lambda_{l}^n = \lambda_{l}^{n,j}$. 
Note that problem \eqref{subproblem1} is a small-scale smooth convex problem and can be solved efficiently through 
the standard convex optimization tool like CVX \cite{CVX}. 
The solutions to problem \eqref{subproblem1} are denoted as $\left\{\xi^n, \bW^n,C_l^n \right\}^{j+1}$.

In the above Step 2, the optimization problem only involves $L$ cache variables $C_l, l \in \mathcal{L}$ and can be formulated as 
\begin{subequations}\label{subproblem2}
 \begin{align} 
\displaystyle \mini_{\left\{C_l\right\}} \hspace{1mm} \quad & \displaystyle  \sum_{l\in\L} \sum_{n\in\N}\left[\lambda_{l}^{n}\left(C_l^n-C_l\right)+\frac{\rho}{2}\left(C_l^n-C_l\right)^2\right] \\
\sbto  \quad &  \displaystyle \sum_{l\in\K} C_l \leq C,~ 0 \leq C_l\leq F,~l\in\L , \\
& \left \vert C_l - C_l(t) \right \vert \leq r(t), ~l\in\L ,
\end{align}
\end{subequations}
which can be reformulated as the following quadratic problem 
\begin{subequations}\label{subproblem22}
 \begin{align} 
\displaystyle \mini_{\left\{C_l\right\}} \quad \hspace{1mm} & \displaystyle  \frac{1}{2}\sum_{l\in\L} \left(C_l - a_l   \right)^2 \\
\sbto  \quad &  \displaystyle \sum_{l\in\L} C_l \leq C,~ 0 \leq C_l\leq F,~l\in\L, \\
& \left \vert C_l - C_l(t) \right \vert \leq r(t), ~l\in\L
\end{align}
\end{subequations}
where $a_l = \frac{\sum_n\left( \rho C_l^{n} + \lambda_{l}^n\right)}{\rho N}$ is a constant with 
$C_l^{n} = C_l^{n, j+1}$ obtained from Step 1 and $\lambda_{l}^n = \lambda_{l}^{n, j}$ obtained 
from the previous iteration. 

With the reformulated problem \eqref{subproblem22}, it is easy to see that 
the optimal $C_l$ admits a closed-form solution given by 
$$C_l^{j+1}=\left[a_l-\mu\right]_{\underline{\theta}_l}^{\bar{\theta}_l},~l\in\L,$$
where $$\underline{\theta}_l = \max \left\{ C_l(t) - r(t), 0 \right\}, ~ 
\bar{\theta}_l = \min \left\{ C_l(t) + r(t), F  \right\},$$
and $\mu$ is the solution to 
$$\sum_{l=1}^L \left[a_l-\mu\right]_{\underline{\theta}_l}^{\bar{\theta}_l} =C$$
conditioned on 
$\sum_{l=1}^L a_l > C$; otherwise $\mu = 0$. 
The desired $\mu$ can be found within $\mathcal{O} \left( L \log_2(L) \right)$ operations. 

In the above proposed ADMM algorithm, 
we introduce a set of auxiliary variables for problem \eqref{prob:re_sample2}, 
which is then optimized over two separate blocks of variables $\left\{\xi^n, \bW^n, C_l^n\right\}$ and 
$\left\{ C_l  \right\}$. 
In \cite[Section 3.2]{ADMM} and \cite[Proposition 15]{AugmentedLA}, the convergence guarantee of such a two-block ADMM algorithm is established 
based on two sufficient conditions: one is that the objective function is closed, 
proper, and convex; the other is that the Lagrangian has at least one 
saddle point. 
It is simple to check that both of the conditions hold for the reformulated problem \eqref{prob:re_sample3}, 
which is equivalent to problem \eqref{prob:re_sample2}. 
Hence, the ADMM algorithm developed above converges to the global optimal solution 
of problem \eqref{prob:re_sample2}. 
%However, it is worth noting here again that, due to the linearization in \eqref{eq:linear}, 
%the overall Algorithm~\ref{alg:cache_alloc} only converges 
%to a stationary point of the nonconvex problem \eqref{prob:re_sample}, or equivalently 
%the original problem \eqref{prob:mbf} when the sample size $N$ is large enough. 

\section{The ADMM Approach to Solve Problem \eqref{prob:re_sample2_multi}}\label{apdx:b}

Similar to problem \eqref{prob:re_sample3}, we first introduce a set of consensus constraints 
$C_{lk}^n = C_{lk},~l\in\L,~k\in\K,~n\in\N$ for problem \eqref{prob:re_sample2_multi} and 
replace the variable $C_{lk}$ in \eqref{lineart_multi} with $C_{lk}^n$. 
Then, the partial augmented Lagrangian of problem \eqref{prob:re_sample2_multi} can be written as 
\begin{align}\label{AL_multi}
 & \mathcal{L}_{\rho}\left( \xi_k^n, \bW_k^n,C_{lk}^n, C_{lk}; \lambda_{lk}^n \right)   = 
\sum_{k=1}^K  \sum_{n=1}^N p_k \frac{1}{\xi_k^n} + \\ \nonumber 
 & ~~~ \sum_{k\in\K}\sum_{l\in\mathcal{L}}\sum_{n\in\N} \left[\lambda_{lk}^n\left(C_{lk}^n-C_{lk}\right)+
\frac{\rho}{2}\left(C_{lk}^n-C_{lk}\right)^2\right],
\end{align}
where $\lambda_{lk}^n$ is the Lagrange multiplier corresponding to the consensus constraint $C_{lk}^n = C_{lk}$. 

As in the three steps listed in Appendix~\ref{apdx:a}, the first step at iteration $j+1$ 
of the ADMM approach to solve 
problem \eqref{prob:re_sample2_multi} is to fix $\left\{ C_{lk}, \lambda_{lk}^n  \right\}$ 
as $C_{lk} = C_{lk}^j, \lambda_{lk}^n = \lambda_{lk}^{n,j} $ obtained from the $j$-th iteration and solve 
for $\left\{\xi_k^n, \bW_k^n, C_{lk}^n\right\}$ by minimizing the Lagrangian \eqref{AL_multi}, 
which is decoupled among each pair of sample channel realization and file index $(n, k),~n\in\N,~k\in\K$. 
The subproblem to be solved in the first step is formulated as follows: 
\begin{subequations}\label{subproblem1_multi}
 \begin{align} 
\displaystyle \mini_{\left\{ \xi_k^n, \bW_k^n,C_{lk}^n\right\}} \quad &  \displaystyle  \frac{p_k}{\xi_k^n} +  
\sum_{l\in\L} \left[\lambda_{lk}^{n}\left(C_{lk}^n-C_{lk}\right)+\frac{\rho}{2}\left(C_{lk}^n-C_{lk}\right)^2\right]\\
 \sbto \hspace{1mm} \quad &  \log \left( 1+\frac{\Tr\left(\bH_{lk}^n\bW_k^n\right)}{\sigma^2} \right) \geq \xi_k^{n}(t) \left( F - C_{lk}^n \right) \nonumber \\  
& \hspace{0.1cm} + (F-C_{lk}(t)) \left(\xi_k^n - \xi_k^{n}(t)\right), ~l\in\L, \\ 
& \Tr(\bW_k^n) \leq P, ~\bW_k^n \succeq\mathbf{0}, \\ 
& \left \vert \xi_k^n - \xi_k^{n}(t) \right \vert \leq r(t) ~. % ~n\in\N 
\end{align}
\end{subequations}
The solutions to the above subproblem \eqref{subproblem1_multi} are denoted as 
$\left\{\xi_k^n, \bW_k^n,C_{lk}^n \right\}^{j+1}$. 

%Note that in the above problem \eqref{subproblem1_multi}, both $C_{lk}$ and $\lambda_{lk}^n$ are fixed constants obtained 
%from the previous iteration of the ADMM procedure. 

In the second step, variables $C_{lk},~l\in\L,~k\in\K$ are updated by minimizing the Lagrangian \eqref{AL_multi} 
under the total cache constraint 
with fixed $C_{lk}^n = C_{lk}^{n, j+1}$ obtained 
from solving problem \eqref{subproblem1_multi} as well as 
fixed $ \lambda_{lk}^n  = \lambda_{lk}^{n,j}$ from the previous iteration. 
The subproblem in the second step can be formulated as 
\begin{subequations}\label{subproblem22_multi}
 \begin{align} 
\displaystyle \mini_{\left\{C_{lk}\right\}} \quad \hspace{1mm} & \displaystyle  
\frac{1}{2}\sum_{l\in\L} \sum_{k\in\K}\left(C_{lk} - b_{lk}   \right)^2 \\
\sbto  \quad &  \displaystyle \sum_{l, k} C_{lk} \leq C,~0 \leq C_{lk} \leq F,~l  \in \mathcal{L}, ~ k \in \mathcal{K}, \\
& \left \vert C_{lk} - C_{lk}(t) \right \vert \leq r(t), ~l\in\mathcal{L},~k\in\K, \label{const:multi_cache_radius}
\end{align}
\end{subequations}
where $b_{lk} = \frac{\sum_n\left( \rho C_{lk}^{n} + \lambda_{lk}^n\right)}{\rho N},~l\in\L,~k\in\K$ are constants. 
The solution to problem \eqref{subproblem22_multi} can be written as 
$$C_{lk}^{j+1}=\left[b_{lk}-\nu\right]_{\underline{\theta}_{lk}}^{\bar{\theta}_{lk}},~l\in\L,~k\in\K,$$
where $$ \underline{\theta}_{lk} = \max \left\{ C_{lk}(t) - r(t), 0 \right\},~ 
\bar{\theta}_{lk} = \min \left\{ C_{lk}(t) + r(t), F \right\},$$
and $\nu$ is the solution to 
$$\sum_{l=1}^L \sum_{k=1}^K \left[b_{lk}-\nu\right]_{\underline{\theta}_{lk}}^{\bar{\theta}_{lk}} =C$$
if 
$\sum_{l=1}^L\sum_{k=1}^K b_{lk} > C$; otherwise $\nu = 0$. 
The desired $\nu$ can be found within $\mathcal{O} \left( LK \log_2(LK) \right)$ operations.

In the last step, we update the Lagrange multiplier $\lambda_{lk}^n$ as 
\begin{equation*} 
	\lambda_{lk}^{n, j+1} := \lambda_{lk}^{n,j} + 	\rho\left( C_{lk}^{n, j+1} - C_{lk}^{j+1}\right),~\forall~l,~k,~n. 
\end{equation*}

%\section*{Acknowledgment}



%\bibliographystyle{IEEEtran}
%\begin{thebibliography}{1}
  %0.5em minus 0.4em\relax Harlow, England: Addison-Wesley, 1999.
%
%\end{thebibliography}

\bibliographystyle{IEEEtran}
%\bibliography{strings,refs}

\bibliography{IEEEabrv,myref}

\begin{thebibliography}{10}
\providecommand{\url}[1]{#1}
\csname url@samestyle\endcsname
\providecommand{\newblock}{\relax}
\providecommand{\bibinfo}[2]{#2}
\providecommand{\BIBentrySTDinterwordspacing}{\spaceskip=0pt\relax}
\providecommand{\BIBentryALTinterwordstretchfactor}{4}
\providecommand{\BIBentryALTinterwordspacing}{\spaceskip=\fontdimen2\font plus
\BIBentryALTinterwordstretchfactor\fontdimen3\font minus
  \fontdimen4\font\relax}
\providecommand{\BIBforeignlanguage}[2]{{%
\expandafter\ifx\csname l@#1\endcsname\relax
\typeout{** WARNING: IEEEtran.bst: No hyphenation pattern has been}%
\typeout{** loaded for the language `#1'. Using the pattern for}%
\typeout{** the default language instead.}%
\else
\language=\csname l@#1\endcsname
\fi
#2}}
\providecommand{\BIBdecl}{\relax}
\BIBdecl

\bibitem{binbin18ICASSP}
B.~Dai, W.~Yu, and Y.-F. Liu, ``Cloud radio access network with optimized
  base-station caching,'' in \emph{Proc. {IEEE} Int. Conf. Acoust., Speech, and
  Signal Process. (ICASSP)}, Apr. 2018.

\bibitem{Rost14}
P.~Rost, C.~Bernardos, A.~Domenico, M.~Girolamo, M.~Lalam, A.~Maeder,
  D.~Sabella, and D.~W\"{u}bben, ``Cloud technologies for flexible {5G} radio
  access networks,'' \emph{{IEEE} Commun. Mag.}, vol.~52, no.~5, pp. 68--76,
  May 2014.

\bibitem{Simeone16}
O.~Simeone, A.~Maeder, M.~Peng, O.~Sahin, and W.~Yu, ``Cloud radio access
  network: Virtualizing wireless access for dense heterogeneous systems,''
  \emph{J. Commun. Netw.}, vol.~18, no.~2, pp. 135--149, Apr. 2016.

\bibitem{CRAN_book}
T.~Q.~S. Quek, M.~Peng, O.~Simeone, and W.~Yu, Eds., \emph{Cloud Radio Access
  Networks: Principles, Technologies, and Applications}.\hskip 1em plus 0.5em
  minus 0.4em\relax Cambridge University Press, 2017.

\bibitem{simeone2009}
O.~Simeone, O.~Somekh, H.~V. Poor, and S.~Shamai~(Shitz), ``Downlink multicell
  processing with limited-backhaul capacity,'' \emph{EURASIP J. Adv. Signal
  Process.}, vol. 2009, no.~1, pp. 1--10, Feb. 2009.

\bibitem{marsch2008}
P.~Marsch and G.~Fettweis, ``On base station cooperation schemes for downlink
  network {MIMO} under a constrained backhaul,'' in \emph{Proc. {IEEE} Global
  Commun. Conf. (Globecom)}, Nov. 2008, pp. 1--6.

\bibitem{Gesbert11}
R.~Zakhour and D.~Gesbert, ``Optimized data sharing in multicell {MIMO} with
  finite backhaul capacity,'' \emph{{IEEE} Trans. Signal Process.}, vol.~59,
  no.~12, pp. 6102--6111, Dec. 2011.

\bibitem{BinbinSparseBFJnal}
B.~Dai and W.~Yu, ``Sparse beamforming and user-centric clustering for downlink
  cloud radio access network,'' \emph{IEEE Access, Special Issue on Recent
  Advances in Cloud Radio Access Networks}, vol.~2, pp. 1326--1339, 2014.

\bibitem{Park13}
S.-H. Park, O.~Simeone, O.~Sahin, and S.~Shamai, ``Joint precoding and
  multivariate backhaul compression for the downlink of cloud radio access
  networks,'' \emph{{IEEE} Trans. Signal Process.}, vol.~61, no.~22, pp.
  5646--5658, Nov. 2013.

\bibitem{PratikEUSIPCO}
P.~Patil, B.~Dai, and W.~Yu, ``Performance comparison of data-sharing and
  compression strategies for cloud radio access networks,'' in \emph{Proc.
  European Signal Process. Conf. (EUSIPCO)}, Aug. 2015, pp. 2456--2460.

\bibitem{Dai16}
B.~Dai and W.~Yu, ``Energy efficiency of downlink transmission strategies for
  cloud radio access networks,'' \emph{{IEEE} J. Sel. Areas Commun.}, vol.~34,
  no.~4, pp. 1037--1050, Apr. 2016.

\bibitem{FLC}
M.~A. Maddah-Ali and U.~Niesen, ``Fundamental limits of caching,'' \emph{{IEEE}
  Trans. Inf. Theory}, vol.~60, no.~5, pp. 2856--2867, May 2014.

\bibitem{Sezgin16}
Y.~Ugur, Z.~H. Awan, and A.~Sezgin, ``Cloud radio access networks with coded
  caching,'' in \emph{Proc. 20th Int. ITG Workshop on Smart Antennas}, Mar.
  2016, pp. 1--5.

\bibitem{Tao16}
M.~Tao, E.~Chen, H.~Zhou, and W.~Yu, ``Content-centric sparse multicast
  beamforming for cache-enabled cloud {RAN},'' \emph{{IEEE} Trans. Wireless
  Commun.}, vol.~15, no.~9, pp. 6118--6131, Sept. 2016.

\bibitem{Park16}
S.~H. Park, O.~Simeone, and S.~S. Shitz, ``Joint optimization of cloud and edge
  processing for fog radio access networks,'' \emph{{IEEE} Trans. Wireless
  Commun.}, vol.~15, no.~11, pp. 7621--7632, Nov. 2016.

\bibitem{Simeone17}
A.~Sengupta, R.~Tandon, and O.~Simeone, ``Fog-aided wireless networks for
  content delivery: Fundamental latency tradeoffs,'' \emph{{IEEE} Trans. Inf.
  Theory}, vol.~63, no.~10, pp. 6650--6678, Oct. 2017.

\bibitem{Patil14}
P.~Patil and W.~Yu, ``Hybrid compression and message-sharing strategy for the
  downlink cloud radio-access network,'' in \emph{Proc. Inf. Theory and
  Applicat. Workshop (ITA)}, Feb. 2014, pp. 1--6.

\bibitem{Gitzenis13}
S.~Gitzenis, G.~S. Paschos, and L.~Tassiulas, ``Asymptotic laws for joint
  content replication and delivery in wireless networks,'' \emph{{IEEE} Trans.
  Inf. Theory}, vol.~59, no.~5, pp. 2760--2776, May 2013.

\bibitem{Shanmugam13}
K.~Shanmugam, N.~Golrezaei, A.~G. Dimakis, A.~F. Molisch, and G.~Caire,
  ``Femtocaching: Wireless content delivery through distributed caching
  helpers,'' \emph{{IEEE} Trans. Inf. Theory}, vol.~59, no.~12, pp. 8402--8413,
  Dec. 2013.

\bibitem{Debbah15}
E.~Ba{\c{s}}tu{\v{g}}, M.~Bennis, M.~Kountouris, and M.~Debbah, ``Cache-enabled
  small cell networks: {Modeling} and tradeoffs,'' \emph{EURASIP J. Wireless
  Commun. Net.}, vol. 2015, no.~1, p.~41, Feb. 2015.

\bibitem{Cui16}
Y.~Cui, D.~Jiang, and Y.~Wu, ``Analysis and optimization of caching and
  multicasting in large-scale cache-enabled wireless networks,'' \emph{{IEEE}
  Trans. Wireless Commun.}, vol.~15, no.~7, pp. 5101--5112, Jul. 2016.

\bibitem{Tao17}
X.~Xu and M.~Tao, ``Modeling, analysis, and optimization of coded caching in
  small-cell networks,'' \emph{{IEEE} Trans. Commun.}, vol.~65, no.~8, pp.
  3415--3428, Aug. 2017.

\bibitem{BidokhtiWT16}
\BIBentryALTinterwordspacing
S.~S. Bidokhti, M.~A. Wigger, and R.~Timo, ``Noisy broadcast networks with
  receiver caching,'' 2016. [Online]. Available:
  \url{http://arxiv.org/abs/1605.02317}
\BIBentrySTDinterwordspacing

\bibitem{BidokhtiWT17}
\BIBentryALTinterwordspacing
S.~S. Bidokhti, M.~A. Wigger, and A.~Yener, ``Benefits of cache assignment on
  degraded broadcast channels,'' 2017. [Online]. Available:
  \url{http://arxiv.org/abs/1702.08044}
\BIBentrySTDinterwordspacing

\bibitem{stochastic_book}
J.~R. Birge and F.~Louveaux, \emph{Introduction to Stochastic Programming},
  2nd~ed.\hskip 1em plus 0.5em minus 0.4em\relax Springer, 2011.

\bibitem{ADMM}
S.~Boyd, N.~Parikh, E.~Chu, B.~Peleato, and J.~Eckstein, ``Distributed
  optimization and statistical learning via the alternating direction method of
  multipliers,'' \emph{Foundations and Trends® in Machine Learning}, vol.~3,
  no.~1, pp. 1--122, 2011.

\bibitem{Liu17}
Y.-F. Liu and W.~Yu, ``Wireless multicast for cloud radio access network with
  heterogeneous backhaul,'' in \emph{Proc. IEEE 51st Asilomar Conf. Signals,
  Syst. and Comput.}, Oct 2017, pp. 531--535.

\bibitem{NIT12}
A.~E. Gamal and Y.-H. Kim, \emph{Network Information Theory}.\hskip 1em plus
  0.5em minus 0.4em\relax Cambridge University Press, 2012.

\bibitem{luo_davidson}
N.~D. Sidiropoulos, T.~N. Davidson, and Z.-Q. Luo, ``Transmit beamforming for
  physical-layer multicasting,'' \emph{{IEEE} Trans. Signal Process.}, vol.~54,
  no.~6, pp. 2239--2251, Jun. 2006.

\bibitem{Lu17}
C.~Lu and Y.-F. Liu, ``An efficient global algorithm for single-group multicast
  beamforming,'' \emph{{IEEE} Trans. Signal Process.}, vol.~65, no.~14, pp.
  3761--3774, Jul. 2017.

\bibitem{CVX}
\BIBentryALTinterwordspacing
M.~Grant and S.~Boyd, ``{CVX}: Matlab software for disciplined convex
  programming, version 2.0 beta,'' Sept. 2013. [Online]. Available:
  \url{http://cvxr.com/cvx}
\BIBentrySTDinterwordspacing

\bibitem{Xu17}
F.~Xu, M.~Tao, and K.~Liu, ``Fundamental tradeoff between storage and latency
  in cache-aided wireless interference networks,'' \emph{{IEEE} Trans. Inf.
  Theory}, vol.~63, no.~11, pp. 7464--7491, Nov. 2017.

\bibitem{Nocedal2006NO}
J.~Nocedal and S.~J. Wright, \emph{Numerical Optimization}, 2nd~ed.\hskip 1em
  plus 0.5em minus 0.4em\relax Springer, 2006.

\bibitem{Yuan1985}
Y.~Yuan, ``Conditions for convergence of trust region algorithms for nonsmooth
  optimization,'' \emph{Math. Program.}, vol.~31, no.~2, pp. 220--228, Jun.
  1985.

\bibitem{Conn2000}
A.~R. Conn, N.~I.~M. Gould, and P.~L. Toint, \emph{Trust-region Methods}.\hskip
  1em plus 0.5em minus 0.4em\relax Society for Industrial and Applied
  Mathematics, 2000.

\bibitem{Lozano07}
A.~Lozano, ``Long-term transmit beamforming for wireless multicasting,'' in
  \emph{Proc. {IEEE} Int. Conf. Acoust., Speech, and Signal Process. (ICASSP)},
  vol.~3, Apr. 2007, pp. 417--420.

\bibitem{Zink09}
M.~Zink, K.~Suh, Y.~Gu, and J.~Kurose, ``Characteristics of {YouTube} network
  traffic at a campus network – measurements, models, and implications,''
  \emph{Comput. Netw.}, vol.~53, no.~4, pp. 501--514, 2009.

\bibitem{AugmentedLA}
J.~Eckstein and W.~Yao, ``Augmented lagrangian and alternating direction
  methods for convex optimization: A tutorial and some illustrative
  computational results,'' \emph{RUTCOR Research Reports}, vol.~32, 2012.

\end{thebibliography}

\end{document}